\documentclass[usenatbib]{mn2e}
\usepackage{apjfonts}
\usepackage{epsfig}
\usepackage{amssymb}
\usepackage{amsmath}
\usepackage{fixltx2e} 
\usepackage{ctable}
\usepackage{lscape}
\usepackage{url}

\newcommand{\mergercalcurl}{\url{http://www.cfa.harvard.edu/~phopkins/Site/mergercalc.html}}
\newcommand{\qlfcalcurl}{\url{http://www.cfa.harvard.edu/~phopkins/Site/qlf.html}}
\newcommand\plotone[1]
 {\centering \leavevmode \includegraphics[width={0.99\columnwidth}]{#1}}
\newcommand{\plotside}[1]
 {\centering \leavevmode \includegraphics[width={0.95\textwidth}]{#1}}
\newcommand{\plotsidesmall}[1]
 {\centering \leavevmode \includegraphics[width={0.85\textwidth}]{#1}}
\newcommand{\acknowledgments}{\begin{small}\section*{Acknowledgments}\end{small}}
\newcommand\altaffilmark[1]{$^{#1}$}
\newcommand\altaffiltext[1]{$^{#1}$}
\newcommand{\etal}{et al.}

\newcommand{\msun}{M_{\sun}}
\newcommand{\lsun}{L_{\sun}}

\title[IR LFs From Cosmological Models]{Mergers, AGN, and ``Normal'' Galaxies: Contributions to 
the Distribution of Star Formation Rates and Infrared Luminosity Functions}

\author[Hopkins \etal]{
\parbox[t]{\textwidth}{ 
Philip F.~Hopkins,\thanks{E-mail:phopkins@astro.berkeley.edu}\altaffilmark{1}, 
Joshua D.~Younger\altaffilmark{2}, 
Christopher C.~Hayward\altaffilmark{3}, \\
Desika Narayanan\altaffilmark{3}, 
\&\ Lars Hernquist\altaffilmark{3}
}
\vspace*{6pt} \\
\altaffiltext{1}{Department of Astronomy and Theoretical 
Astrophysics Center, University of California Berkeley, Berkeley, CA 94720} \\
\altaffiltext{2}{Hubble Fellow, Institute for Advanced Study, Einstein Drive,
Princeton, NJ 08540} \\ 
\altaffiltext{3}{Harvard-Smithsonian Center for Astrophysics, 60 
Garden Street, Cambridge, MA 02138, USA}  
}

\date{Submitted to MNRAS, September 17, 2009}
\begin{document}
\maketitle
\label{firstpage}

\begin{abstract}
We use a novel method to predict the contribution of normal star-forming
galaxies, merger-induced bursts, and obscured AGN, to IR luminosity functions
(LFs) and global SFR densities. We use empirical halo occupation constraints to
populate halos with galaxies and determine the distribution of normal and
merging galaxies. Each system can then be associated with high-resolution
hydrodynamic simulations. We predict the distribution of observed luminosities
and SFRs, from different galaxy classes, as a function of redshift from $z=0-6$.
We provide fitting functions for the predicted LFs, quantify the uncertainties,
and compare with observations. At all redshifts, `normal' galaxies dominate the
LF at moderate luminosities $\sim L_{\ast}$ (the `knee'). Merger-induced bursts increasingly
dominate at $L\gg L_{\ast}$; at the most extreme luminosities, AGN are important. However,
all populations increase in luminosity at higher redshifts, owing to increasing
gas fractions. Thus the `transition luminosity' between normal and merger-dominated sources
increases from the LIRG-ULIRG threshold at $z\sim0$ to bright Hyper-LIRG thresholds
at $z\sim2$. The transition to dominance by obscured AGN evolves similarly, at factor
of several higher $L_{\rm IR}$. At all redshifts, non-merging systems dominate the total
luminosity/SFR density, with merger-induced bursts constituting $\sim5-10\%$ and AGN
$\sim1-5\%$. Bursts contribute little to scatter in the SFR-stellar mass relation. In
fact, many systems identified as `ongoing' mergers will be forming stars in
their `normal' (non-burst) mode. Counting this as `merger-induced' star
formation leads to a stronger apparent redshift evolution in the contribution of
mergers to the SFR density. We quantify how the evolution in LFs depends on evolution in galaxy 
gas fractions, merger rates, and possible evolution in the Schmidt-Kennicutt relation. 
We discuss areas where more detailed study, with full radiative transfer treatment of 
complex three-dimensional clumpy geometries in mixed AGN-star forming systems, is necessary. 
\end{abstract}

\begin{keywords}
galaxies: formation --- galaxies: evolution --- galaxies: active --- 
star formation: general --- cosmology: theory
\end{keywords}

\section{Introduction}
\label{sec:intro}

Understanding the global star-formation history of the Universe remains 
an important unresolved goal in cosmology. 
In recent years, observations of the properties of galaxies in 
the infrared, at redshifts $z=0-3$, have begun to shed light on 
this history, but have also revealed a number of intriguing questions. 

Of particular interest are the roles of mergers and AGN in driving 
star formation and/or the infrared luminosities of massive systems. 
A wide range of observed phenomena support the view that gas-rich\footnote{
By ``gas,'' we refer specifically to cold, star-forming 
gas in galaxy disks, as opposed to hot, virialized gas.}
mergers are important to galaxy evolution; but it is less clear what their 
role is in the global star formation process and buildup of stellar mass 
in the Universe. 
In the local Universe, the population of star-forming galaxies appears 
to transition from ``quiescent'' (non-disturbed)\footnote{In this paper, 
we use the term ``quiescent'' to refer to star-forming systems that are not 
strongly disturbed in e.g.\ major mergers and are forming stars in similar 
fashion to most ``normal'' disks. We do {\em not} mean 
non-star forming systems, as the term is used in some 
literature.} disks -- which dominate the 
{\em total} star formation rate/IR luminosity density -- at the luminous 
infrared galaxy (LIRG) threshold $10^{11}\,\lsun$ ($\dot{M}_{\ast}\sim 10-20\,\msun\,{\rm yr^{-1}}$)  
to clearly merging, violently disturbed systems at a few times this luminosity. The most intense
starbursts at $z=0$, ultraluminous infrared galaxies (ULIRGs; $L_{\rm IR}>10^{12}\,\lsun$), are 
invariably
associated with mergers \citep[e.g.][]{joseph85,sanders96:ulirgs.mergers,
evans:ulirgs.are.mergers}, with
dense gas in their centers providing material to feed black hole (BH)
growth and to boost the concentration and central phase space density
of merging spirals to match those of ellipticals
\citep{hernquist:phasespace,robertson:fp}. 
Various studies have shown that the mass involved in these starburst events 
is critical to explain the relations between spirals, mergers, and ellipticals, 
and has a dramatic impact on the properties of merger 
remnants \citep[e.g.,][]{LakeDressler86,Doyon94,ShierFischer98,James99,
Genzel01,tacconi:ulirgs.sb.profiles,dasyra:mass.ratio.conditions,dasyra:pg.qso.dynamics,
rj:profiles,rothberg.joseph:kinematics,hopkins:cusps.ell,hopkins:cores}. 

At high redshifts, the role of mergers is less clear. 
It is clear that LIRGs and ULIRGs increase in relative importance with 
redshift, with LIRGs dominating the star formation rate/IR luminosity 
densities at $z\sim1$ and ULIRGs dominating at $z\sim2$ 
\citep[e.g.][]{lefloch:ir.lfs,perezgonzalez:ir.lfs,caputi:ir.lfs,magnelli:z1.ir.lfs}.
This, together with the fact that merger rates are expected 
and observed to increase with redshift \citep[by a factor $\sim10$ from 
$z=0-2$; see e.g.][and references therein]{hopkins:merger.rates} 
has led to speculation that the merger rate evolution may in fact 
drive the observed evolution in the cosmic SFR density, which rises 
rapidly from $z\sim0-2$ and then turns over, declining more slowly 
\citep[e.g.][and references therein]{hopkinsbeacom:sfh}. 

However, many LIRGs at $z\sim1$, and potentially 
ULIRGs at $z\sim2$, appear to be ``normal'' galaxies, 
without dramatic morphological disturbances associated with the 
local starburst population or large apparent AGN contributions 
\citep{yan:z2.sf.seds,sajina:pah.qso.vs.sf,
dey:2008.dog.population,melbourne:2008.dog.morph.smooth,
dasyra:highz.ulirg.imaging.not.major}. 
At the same time, even more luminous systems 
appear, including large populations of Hyper-LIRG (HyLIRG; 
$L_{\rm IR}>10^{13}\,\lsun$) and bright sub-millimeter galaxies 
\citep[e.g.][]{chapman:submm.lfs,younger:highz.smgs,
younger:sma.hylirg.obs,casey:highz.ulirg.pops}. 
These systems exhibit many of the traits more commonly 
associated with merger-driven starbursts, including morphological 
disturbances, and may be linked to the emergence of 
massive, quenched (non star-forming), compact ellipticals 
at times as early as $z\sim2-4$ 
\citep{papovich:highz.sb.gal.timescales,
younger:smg.sizes,tacconi:smg.maximal.sb.sizes,
schinnerer:submm.merger.w.compact.mol.gas,
chapman:submm.halo.clustering,tacconi:smg.mgr.lifetime.to.quiescent}. 
But reproducing their abundance 
and luminosities remains a challenge for current models of 
galaxy formation \citep{baugh:sam,
swinbank:smg.counts.vs.durham,
narayanan:smg.modeling,
younger:warm.ulirg.evol}.

In a related vein, observations of a tight correlation between the 
masses of super-massive BHs and their host spheroid properties 
\citep{Gebhardt00,FM00,magorrian,novak:scatter,hopkins:bhfp.obs}
suggest a tight coupling between BH growth 
and star formation, perhaps in particular to the mergers 
believed to drive the formation
of the most massive bulges. 
Considering the energy output required to form the BH 
population \citep[e.g.][]{soltan82}, or the observed bolometric 
quasar energy density as a function of redshift 
\citep[see][and references therein]{hopkins:bol.qlf}, 
it is clear that the {\em bolometric} output of quasars and AGN 
is at least roughly comparable to the total infrared luminosity density of 
the Universe at most redshifts ($z\sim1-3$) -- although 
the measurements above suggest it is still a factor $\sim2-3$ lower. 
Some recent observations have suggested that the population of 
very luminous, highly obscured (Compton-thick) quasars 
may be considerably larger than previous estimates, 
in which case the heavily-obscured AGN population could 
represent a large fraction of the total IR luminosity density 
at high redshifts \citep{hickox:bootes.obscured.agn,daddi:2007.high.compton.thick.pops,
treister:compton.thick.fractions}. 
This would have dramatic implications not just for BH populations 
and e.g.\ the implied radiative efficiencies of BH accretion, but also 
for the implied total star formation rate density. 
Some apparent discrepancies between e.g.\ the total mass density 
observed in old stars and the implied star formation rate density 
have been cited as possible evidence of a time-dependent 
stellar initial mass function (IMF); but a rising contribution from 
obscured AGN at high redshifts could mimic this effect 
\citep{hopkinsbeacom:sfh,dave:imf.evol}. 

In particular, there are long-standing questions of what powers 
the most luminous infrared sources, for example, ULIRGs and 
sub-millimeter galaxies. 
This debate extends to the discovery of these objects 
\citep[see e.g.][]{soifer84b,soifer:iras,scoville86,
sargent87,sanders88:warm.ulirgs,
solomon.downes:ulirg.ism}, 
and has persisted despite the addition of millimeter spectroscopy 
and observations in a large number of independent 
wavebands \citep[for a review of the debate, see both][]{sanders:agn.vs.sf.in.ulirgs,
joseph:sb.vs.agn.power.ulirgs}. 
Although some evidence suggests that they are primarily powered by 
star formation \citep{farrah:qso.vs.sf.sed.fitting,
lutz:pah.qso.vs.sf.local,sajina:pah.qso.vs.sf,
pope:2008.pah.agn.dont.dominate.smgs,pope:2008.dog.sfr.properties,
watabe:highz.ulirg.sb.vs.agn,nardini:2009.agn.vs.sb.contrib.in.ulirgs}, 
the constraints and correlations typically invoked have inherent 
factor $\sim2$ uncertainties, and thus could easily accommodate comparable 
power input from star formation and AGN. 
Moreover, a sufficiently obscured AGN, in a medium with the right optical 
depth properties, is indistinguishable from star formation by the usual indicators 
(e.g.\ PAH strengths, emission region sizes, or any other infrared spectral or 
morphological criteria). Hence even at 
$z=0$, debate surrounds the power source of many bright infrared systems, 
and there exist a number of examples of systems 
classified as ``star formation dominated'' by all of these metrics that 
later revealed Compton-thick AGN whose longer-wavelength emission 
has been fully re-processed, even into ``cool'' dust 
\citep[see e.g.][and references therein]{alexander:2008.compton.thick.z2.qsos}. 
In a bolometric sense, the most luminous galaxies observed 
(with $L\gtrsim 10^{14}\,\lsun$ or $\gg 10^{47}\,{\rm erg\,s^{-1}}$) 
are the most luminous quasars; although the contribution of these systems to the 
infrared remains highly uncertain. 
This may be important for resolving the theoretical difficulties in modeling these 
bright systems. 

There has been important theoretical progress in modeling 
these processes in an {\em a priori} manner \citep[see e.g.][]{baugh:sam,
hopkins:groups.qso,narayanan:smg.modeling,younger:warm.ulirg.evol}. 
However, two basic 
limitations remain. In direct cosmological hydrodynamic simulations,
as well as semi-analytic models of galaxy formation, it is well known that 
it remains challenging to accurately reproduce global quantities such as 
the galaxy mass function and the distribution of sizes, gas fractions, and 
hence star formation rates, especially the distributions of star-forming 
gas and their relations to whether or not galaxies are ``quenched''
\citep[recently, see e.g.][]{weinmann:group.cat.vs.sam,
maller:sph.merger.rates,kimm:passive.sats.vs.centrals,fontanot:downsizing.vs.sams}. 
This makes it difficult to determine whether discrepancies between such 
models and the observations owe to their treatment of star formation, 
or to discrepancies in these quantities. Moreover, it is difficult to 
disentangle the effects of these different properties on the distribution of 
star formation rates. In addition, for merger-induced starburst and AGN 
activity, although it may be possible to roughly estimate some global quantities 
(e.g.\ the total mass involved in a starburst) from simple analytic motivations or 
low-resolution cosmological simulations (several $\sim$kpc typical), 
it is not straightforward to estimate the chaotic, time-dependent 
behavior of full lightcurves needed to estimate the distribution of time spent at 
different, rapidly varying luminosities without high-resolution simulations 
of individual systems. Since the 
number density of the most bright systems is exponentially declining, 
fluctuations and features in the starburst/AGN fueling history on small 
time and spatial scales ($\Delta t\sim10^{7}\,{\rm yr}$, $R\lesssim 100\,$pc) 
can be critical for correct estimates of their contributions to 
bright populations. 

In this paper, we present theoretical predictions for the 
distribution of galaxy star formation rates and infrared luminosities, 
as a function of galaxy mass and redshift, using a novel methodology 
that can circumvent some of these obstacles. 
We combine a halo-occupation based approach, in which we take 
galaxy properties as fixed from observations at each epoch, 
and then apply rules for the distribution of star formation rates/infrared luminosities 
in ``quiescent'' systems, merger-induced starbursts, and obscured AGN, 
calculated from a large suite of high-resolution hydrodynamic simulations of individual 
galaxies and galaxy mergers. 
We use this to independently estimate the contributions of 
``normal'' galaxies, mergers, and AGN to the luminosity functions. 
The comparisons we make are approximate -- we do not include full 
time-dependent radiative transfer in simulations (the subject of future work, in progress), 
and so focus on integral quantities such as the total IR luminosity and SFR distributions, 
that are less sensitive to issues of e.g.\ the exact dust distribution, 
temperature, and other properties. We also explicitly separate the contributions of 
AGN and star formation, but stress that, in real systems where the two 
are comparable, their additive effects are non-linear, and will require further study. 
We show how adding or removing components of the model taken from 
observations such as e.g.\ the distribution of galaxy sizes and gas fractions 
affects these consequences. We compare to observations of all quantities, 
where available, and find reasonable agreement but with some interesting 
apparent discrepancies at high redshifts. We also show how these 
populations relate to the scatter in star formation rates at fixed galaxy masses, 
and in a global sense to the total star formation rate density, the star formation 
rate density in mergers, and the fraction of the 
inferred star formation rate density which might really be driven by 
obscured AGN activity. Readers interested primarily in 
the comparison of predictions and observations 
may wish to skip directly to \S~\ref{sec:lfs}. 

Throughout, we adopt a $\Omega_{\rm M}=0.3$, $\Omega_{\Lambda}=0.7$,
$h=0.7$ cosmology and a \citet{chabrier:imf} stellar IMF (discussed further below), but 
these choices do not affect our conclusions.

\section{The Model}
\label{sec:model}

\begin{figure*}
    \centering
    \plotsidesmall{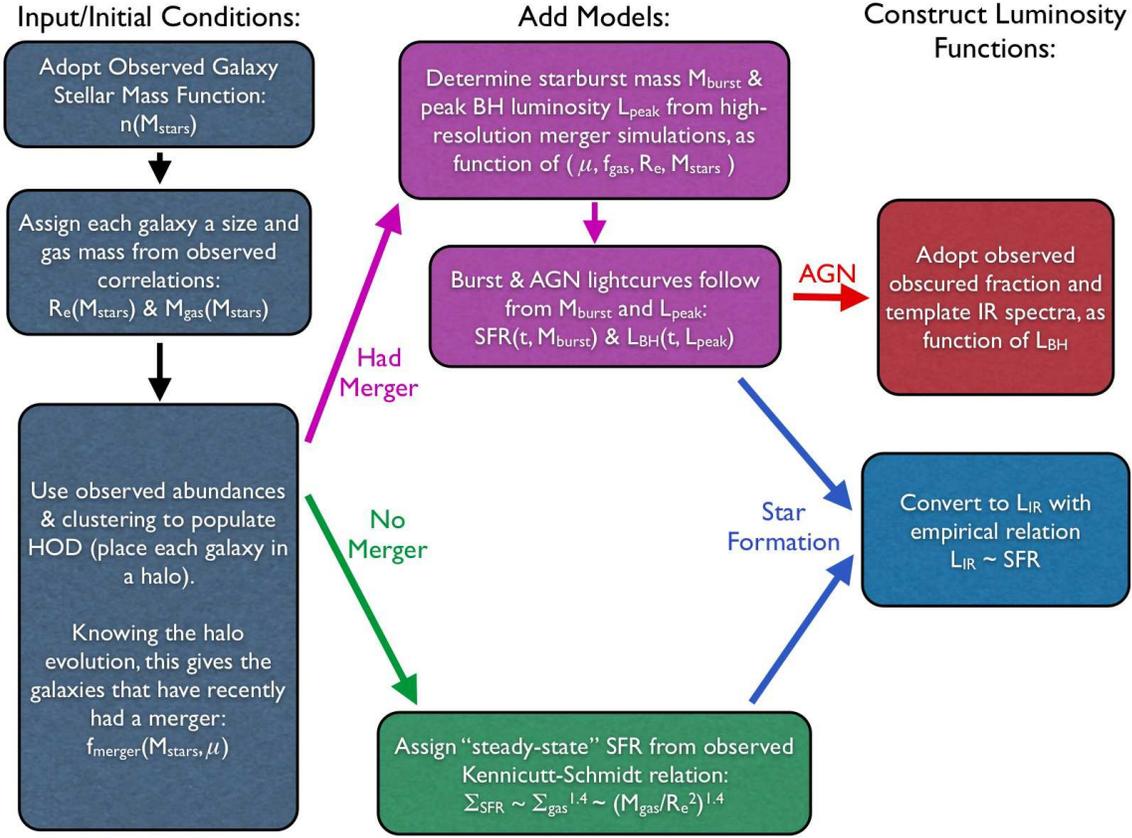}
    \caption{Summary of our model methodology, described in detail in 
    \S~\ref{sec:model}. We begin with a halo occupation model: at a given 
    redshift, the galaxy stellar mass function, and distribution of galaxy sizes 
    and gas fractions are taken from observations. Placing galaxies in halos 
    from dark matter simulations 
    according to their observed abundance and clustering, and evolving these 
    forward in time some short time, we obtain the merger population. 
    Non-merging galaxies are assigned a star formation rate based on their 
    size and gas mass, according to the observed Kennicutt-Schmidt relation; 
    for mergers, the lightcurves of merger-induced starburst and AGN activity 
    are taken from fits to high-resolution galaxy merger simulations (including 
    star formation based on a local Kennicutt-Schmidt law, gas cooling, 
    and feedback from accretion and star formation), as a function 
    of salient galaxy properties. Star formation rates are converted to infrared 
    luminosities with a simple empirical proportionality; AGN bolometric 
    luminosities are corrected to far-IR luminosities based on observed obscured 
    fractions. 
    \label{fig:methodology}}
\end{figure*}

The model used here is a slightly modified version of one that has been discussed 
extensively in a series of papers \citep[most recently][]{hopkins:disk.survival.cosmo,
hopkins:merger.rates}. We summarize the salient properties here. 
Figure~\ref{fig:methodology} provides a simple outline of the model, 
on which we elaborate below.

\subsection{Halo Occupation Constraints: The Initial Galaxy Population from Observations}
\label{sec:model:hod}

At a given redshift, we use the halo occupation distribution (henceforth HOD) formalism to 
construct a mock sample of galaxies. 
Specifically, we begin with the observed galaxy stellar mass function (MF), 
which we take as given. 
Since we are interested specifically in star-forming galaxies, we 
adopt just the galaxy stellar mass function of star forming or 
``blue'' galaxies where available 
(here at all $z<2$); although there may be some trace star formation 
in red galaxies, assuming typical values yields a negligible contribution 
to the bright far-IR and massively star-forming populations. 
At redshifts $z>2$, type-separated MFs are no longer available, 
so we simply adopt the total galaxy MF (i.e.\ assume all systems 
are star-forming); however, the fraction of massive 
galaxies that are ``quenched'' and red has become sufficiently low 
by $z=2$ (and is rapidly falling) that it makes little difference 
(e.g.\ adopting the upper limit -- that the red fraction at all 
masses at $z>2$ is equal to that at $z=2$ -- makes no difference to 
our predictions). 

The uncertainties in the galaxy abundance are one of the 
dominant uncertainties in the model, especially at high redshifts. 
We therefore consider different mass function fits, to represent the 
possible range. At $z=0$, the uncertainties are relatively small; 
we adopt the mass function of star forming galaxies from 
\citet{bell:mfs}. From $z=0-2$, we consider the MFs of 
star-forming galaxies from \citet{arnouts:2007.type.sep.mfs.to.z2} 
and \citet{ilbert:cosmos.morph.mfs}. The range between the 
two is representative of the uncertainties and scatter in a number of 
other calculations, which cover different portions of this 
dynamic range \citep[e.g.][]{bundy:mfs,pannella:mfs,franceschini:mfs,
borch:mfs,fontana:highz.mfs,brown:mf.evolution}. 
At $z>2$, we adopt as bracketing the relevant dynamic range 
the mass functions from \citet{perezgonzalez:mf.compilation} 
and \citet{fontana:highz.mfs}; again, other determinations 
\citep[e.g.][]{marchesini:highz.stellar.mfs,kajisawa:stellar.mf.to.z3} 
lie within this range. 

Given each galaxy and its stellar mass, 
we assign it other properties in accord with observations. 
First, a gas mass. It is well-established that, at fixed stellar mass, 
galaxy gas fractions are higher at high redshifts (see 
compiled references below). Moreover, the 
trend of galaxy gas fractions as a function of stellar mass, 
and their scatter, have been quantified (both directly 
and indirectly) at a range of 
redshifts from $z=0$ to $z=3$. 
We have compiled observations from the available sources, 
spanning this redshift range and a stellar mass range from 
$M_{\ast}\sim 10^{10}-10^{12}\,\msun$ (more than sufficient dynamic 
range for the predictions of interest here), 
specifically from \citet{belldejong:disk.sfh,mcgaugh:tf,
calura:sdss.gas.fracs,shapley:z1.abundances,erb:lbg.gasmasses,
puech:tf.evol,mannucci:z3.gal.gfs.tf,cresci:dynamics.highz.disks,
forsterschreiber:z2.sf.gal.spectroscopy,erb:outflow.inflow.masses}. 
We present these observations in a number of papers \citep{hopkins:groups.qso,
hopkins:groups.ell,hopkins:bhfp.theory,hopkins:disk.survival,
hopkins:r.z.evol} and show that the $z=0$ zero-point 
and evolution with redshift can be well-fitted by the simple functions 
\begin{align}
\nonumber f_{\rm gas}(M_{\ast}\,|\,z=0) &\equiv f_{0}\approx \frac{1}{1+(M_{\ast}/10^{9.15})^{0.4}} \\
f_{\rm gas}(M_{\ast}\,|\,z) &= f_{0}\,
\{ 1 - \tau(z)\,[1-f_{0}^{3/2}] \}^{-2/3} \ ,
\label{eqn:fgas.z}
\end{align}
where $\tau(z)$ is the fractional lookback time to a redshift 
$z$ ($\equiv0$ at $z=0$ and $\equiv1$ at $z\rightarrow \infty$).
The former functional form is motivated by cosmological hydrodynamic 
simulations \citep{keres:hot.halos,keres:cooling.revised}, 
and the latter by the scalings of simple closed-box models 
that obey the \citet{kennicutt98} relation at all times.\footnote{This 
function is presented in \citet{hopkins:cusps.evol}, 
Equation~(2); we note that in the text there is a typo, 
and the equation is written with a $+\tau(z)$. The correct form (above), 
with $-\tau(z)$ was, however, used for the calculations therein.} 
The important quantity is not the precise scaling, but rather, the 
fact that it provides a convenient interpolation formula between 
the observations above. We assume a constant, intrinsic 
$0.25\,$dex scatter about these gas masses at each stellar mass, 
also in agreement with the observations above. 

Next, we assign galaxy spatial sizes, again from observations. 
At $z=0$, the distributions of disk sizes are well-measured; 
at higher redshifts, there is some uncertainty, but observations 
are converging on the conclusion that the star-forming population 
evolves relatively mildly in size with redshift (whereas the 
non-star-forming population evolves more rapidly). From 
$z=0-2$, we find that the compilation of observational results on 
the evolution of the disk size-mass relation from 
\citet{trujillo:size.mass.to.z3,
ravindranath:disk.size.evol,ferguson:disk.size.evol,
barden:disk.size.evol,toft:z2.sizes.vs.sfr,akiyama:lbg.weak.size.evol}, 
and the theoretical models in \citet{somerville:disk.size.evol}, 
can be simply represented as relatively weak power-law evolution 
in disk size at fixed mass
\begin{align}
\nonumber R_{e}(M_{\ast}\,|\,z=0) &\approx 5.28\,{\rm kpc}\,{\Bigl (}\frac{M_{\ast}}{10^{10}\,\msun}{\Bigr )}^{0.25} \\ 
R_{e}(M_{\ast}\,|\,z) & = R_{e}(M_{\ast}\,|\,z=0)\,(1+z)^{-0.6}
\label{eqn:size.z}
\end{align}
where the $z=0$ relation is taken from \citet{shen:size.mass} 
(appropriately normalized for our 
adopted cosmology and IMF), and we assume a 
constant $0.2\,$dex scatter in disk sizes. 
As we will show, our results are not especially sensitive to the adopted size evolution, 
so this is not a major source of uncertainty.

\subsection{``Normal'' Star Formation: Relation to Galaxy Properties}
\label{sec:model:kennicutt}

Our major assumption is the Kennicutt-Schmidt law holds 
at all redshifts, relating the (average) surface density of star formation 
to the average surface gas density 
\begin{equation}
\dot{\Sigma}_{\ast} = 1.3\times10^{-4}\,\msun\,{\rm yr^{-1}\,kpc^{-2}}
\,{\Bigl (}\frac{\Sigma_{\rm gas}}{\msun\,{\rm pc^{-2}}} {\Bigr )}^{n_{K}} 
\label{eqn:kennicutt}
\end{equation}
with the best-fit index $n_{K}\approx1.4$ 
\citep{kennicutt98}. 
The normalization here is corrected for our assumed \citet{chabrier:imf} IMF. 
We will later consider the index and normalization to 
be free, but for now take this relation as fixed. 
Some simple algebra shows that this (assuming no dramatic 
evolution in disk profile shapes) can be written in the global form 
\begin{equation}
\frac{\dot{M}_{\ast}}{\msun\,{\rm yr^{-1}}} = 
1.3\,{\Bigl (}\frac{10^{4}}{\pi}{\Bigr )}^{n_{K}-1}\,
{\Bigl (} \frac{M_{\rm gas}}{10^{10}\,\msun} {\Bigr )}^{n_{K}} \,
{\Bigl (} \frac{R_{e}}{{\rm kpc}} {\Bigr )}^{-2\,(n_{K}-1)}\ .
\label{eqn:kennicutt.global}
\end{equation}
Together with the assumptions above, this defines a 
``steady state'' or ``quiescent'' star formation rate for 
all disks in the model. 

We will show that the resulting SFR distributions agree well 
with those observed for normal galaxies, suggesting that these scalings 
are reasonable. However, we have also checked them against 
direct observations of the median SFR of disc galaxies as a function of 
stellar mass and redshift. These are measured in \citet{noeske:sfh} 
from $z=0-1.2$ and in \citet{papovich:ssfr} at $z\sim2$; 
comparing with the simple predictions from Equation~\ref{eqn:kennicutt.global} 
and the equations above yields reasonably good agreement. 
The combined dependence of $f_{\rm gas}(M_{\ast})$ and 
$R_{e}(M_{\ast})$ means that $\dot{M}_{\ast}(M_{\ast})$ weakly increases 
with $M_{\ast}$ in a roughly power-law like fashion, in good agreement 
with these observations; and the redshift dependence of $f_{\rm gas}$ 
yields a similar increase in the normalization of the SFRs with 
redshift (and by construction, the $z=0$ normalizations are similar). 
In fact, we find that our results for the quiescent disk population 
are completely unchanged (within their uncertainties) 
if we simply adopt a parameterized fit 
to the observations in these papers -- i.e.\ if we bypass all of the 
above assumptions and simply adopt a fit to the observed 
$\dot{M}_{\ast}(M_{\ast}\,|\,z)$ relations. However, this would 
severely limit the dynamic range in redshift and mass to which 
we could robustly apply these models, as well as limiting 
the physical insight gained, and (most of all) would not allow for the 
straightforward predictions for merger and AGN populations.

\subsection{Merger Rates and Resulting Starburst Properties}
\label{sec:model:mergers.sb}

We next require a model for merger rates, in order to 
model merger-induced bursts of star formation and AGN activity. 
The methodology for doing so is described and tested in 
\citet{hopkins:merger.rates}, but we briefly summarize here.\footnote{The 
approximate merger rates from the model presented in \citet{hopkins:merger.rates} 
can also be obtained as a function of galaxy mass, merger mass 
ratio, and gas fraction from the publicly available ``merger rate calculator'' 
script at \mergercalcurl.}
We assign each galaxy to a halo or subhalo in a simple manner following 
the standard halo occupation methodology described in 
\citet{conroy:hod.vs.z}; ensuring, by construction, that 
the galaxy mass function and galaxy clustering 
(as a function of stellar mass, galaxy color, and physical scale) is exactly reproduced. 
At a given instant, then, knowing the halo-halo merger rates as a function of 
e.g.\ halo mass and mass ratio, we can convolve this with the 
determined galaxy masses in each halo \citep[and appropriately correct for 
e.g.\ the dynamical friction time delay between halo-halo and galaxy-galaxy merger, 
following][]{boylankolchin:merger.time.calibration}, 
and obtain the galaxy-galaxy merger rate as a function of 
galaxy mass $M_{\ast}$, redshift $z$, and galaxy-galaxy 
baryonic (or stellar) mass ratio $\mu \equiv M_{2}/M_{1}$ 
(defined always so that $M_{1}>M_{2}$, i.e.\ $0<\mu<1$). 
The halo mass functions and merger rates are adopted from the 
Millenium simulation \citep{springel:millenium,fakhouri:halo.merger.rates}; 
but in \citet{hopkins:merger.rates} we compare this with a wide 
variety of alternative simulations and calculations, as well as 
a number of differences in methodology, and show that these all 
lead to small (factor $<2$) differences in the resulting merger rate. 
In that paper, we also compare this calculation to a large number of 
observational constraints in the redshift range $z=0-2$ (see references 
therein), and show that the two agree well; adopting a parameterized 
fit to the observed major merger rate from most observations yields 
an identical result in our calculation here (but does not have the 
convenience of being easy to extrapolate to arbitrary mass 
ratios $\mu$ or redshifts $z$). Note that, again, we begin from just the 
star-forming galaxy luminosity function -- ``dry'' mergers of quiescent 
systems will not produce interesting starburst or AGN activity. 

In a merger, gravitational torques lead to gas in the disc rapidly losing 
angular momentum to the nearby stars, and falling inwards
\citep{barnes.hernquist.91,barneshernquist96}. 
The rapid increase in the central gas densities drives a massive 
starburst \citep{mihos:starbursts.94,mihos:starbursts.96}. 
Here, we assume that 
every merger induces a starburst and corresponding AGN activity. 
In \citet{hopkins:disk.survival}, as well as a number of other 
studies \citep[e.g.][and references therein]{dimatteo:merger.induced.sb.sims,
cox:massratio.starbursts,hopkins:ang.mom.overview}, 
the resulting total starburst mass/amplitude 
and peak QSO luminosity are quantified as a function of merger 
properties, from a suite of hundreds of high-resolution hydrodynamic 
galaxy merger simulations. These simulations span a range in 
the relevant properties: redshift, merger mass ratio, orbital parameters, 
galaxy structural properties, and gas fractions; and they include 
prescriptions based on the same \citet{kennicutt98} law for 
dynamic star formation, as well as black hole accretion and 
feedback from supernovae and AGN \citep{dimatteo:msigma,springel:red.galaxies}. 
Together this allows for a 
full sampling of the interesting parameter space, and a simple, 
direct parameterization of the resulting burst properties. 

Despite the complex physics involved, it is shown therein 
that the average burst scalings can be represented in analytic 
form, motivated by basic gravitational physics. We adopt the 
full scalings derived therein, but note that the important parameter, 
the {\em total} mass of gas that loses angular momentum and 
participates in the central starburst, scales (to lowest order) 
with the simple relation (after averaging over a random distribution of 
orbital parameters) 
\begin{equation}
M_{\rm burst} \sim M_{\rm gas}\,\mu\,(1-f_{\rm gas})\ .
\end{equation}
The scaling with merger 
mass ratio $\mu$ represents the declining efficiency of angular 
momentum loss in more minor mergers; the scaling 
with $(1-f_{\rm gas})$ comes from the fact that the torques that 
remove angular momentum from gas are primarily {\em internal}, 
from stars in the same galaxy -- a pure gas merger would simply yield 
a new disk, not a compact starburst.\footnote{
The physics 
of these scalings, particularly 
that with gas fraction, is discussed in detail in \citet{hopkins:disk.survival}. 
In short, hydrodynamic torques and pressure forces are negligible, 
and direct torques from e.g.\ the secondary galaxy and halo 
are suppressed by a tidal term $\sim (r/R_{e})^{3}$, 
the short time of close passage (much less than the several dynamical 
times needed to continue strong gas inflows), 
and the fact that they are out-of-resonance with the primary gas disk. 
Moreover these torques are just as likely to increase as decrease 
the gas angular momentum. 
As shown in that paper and earlier
\citep{barneshernquist96,barnes:review}, 
this means that the stellar disk in the same galaxy 
(with fractional mass $(1-f_{\rm gas})$), being in 
direct spatial proximity and resonance, always 
dominates the torques driving angular momentum loss in the gas. 
}
Adopting the simplified scaling above, in fact, yields very similar 
results to the full scaling presented in \citet{hopkins:disk.survival} 
for $M_{\rm burst}$ as a function of $\mu$, $M_{\rm gas}$, 
$f_{\rm gas}$, and orbital parameters (we assume random 
orbital inclinations and parabolic orbits, motivated by cosmological 
simulations). Motivated by the simulations (or e.g.\ allowing for the 
full distribution of orbital parameters), we adopt a constant 
$0.35\,$dex scatter in $M_{\rm burst}$ at fixed galaxy properties 
(with of course the limit $M_{\rm burst}<M_{\rm gas}$). 

In a burst, there is some non-trivial time-dependent lightcurve or 
star formation rate versus time. 
Since we are considering the statistical distribution of 
luminosities, we do not need to know the exact time-dependent form 
of this function; rather, the important quantity is the distribution of times 
spent at different luminosities. Examples of this are 
shown in detail in \citet{hopkins:merger.lfs}, and similar quantities 
are presented in \citet{dimatteo:merger.induced.sb.sims,
cox:massratio.starbursts}. 
We find that, integrated over the history of a burst, this function 
can (on average) be conveniently represented by the simple function 
\begin{equation}
\frac{{\rm d}t}{{\rm d}\log{\dot{M}_{\ast}}} = t_{\rm burst}\,\ln{10}\,
\exp{{\Bigl \{} -\frac{\dot{M}_{\ast}}{M_{\rm burst}/t_{\rm burst}} {\Bigr \}}}\ ,
\label{eqn:tburst}
\end{equation}
where, fitting to the simulations, we find $t_{\rm burst}\approx 0.1\,$Gyr, 
nearly independent of galaxy mass and redshift. This functional 
form is characteristic of a rapid, exponential rise from low 
SFR to a peak in the burst, with a burst lifetime of order $t_{\rm burst}$. 
The constancy and normalization of this lifetime is a simple consequence 
of the observed dynamical times in the central regions of galaxies, 
and the fact that these dynamical times scale weakly or not at all 
with mass and redshift \citep[see e.g.][]{belldejong:tf,
mcgaugh:tf,courteau:disk.scalings}. 
Given some merger rate, and corresponding rate of ``creation'' of 
bursts of a given mass (${\rm d}n(M_{\rm burst})/{\rm d}t$), 
the observed number density of bursts at a given 
$\dot{M}_{\ast}$ is simply given by the convolution of this 
rate with the lifetime above, 
i.e.\ ${\rm d}n(M_{\rm burst})/{\rm d}\log{\dot{M}_{\ast}} 
= ({\rm d}n(M_{\rm burst})/{\rm d}t)\,({\rm d}t/{\rm d}\log{\dot{M}_{\ast}}$)  
\citep[for more explicit details of this methodology, see][]{hopkins:qso.all}. 

Note that the numbers above are somewhat different from those 
presented in \citet{hopkins:merger.lfs}. However, in that paper, 
we were considering the total distribution of star formation rates that would 
be observed over the {\em entire} duration in which a system might be identified 
as a merger or interacting pair. The lifetime of that phase is much longer, 
$\sim1-2$\,Gyr, and (by time and by total mass) most of the star formation 
comes from the ``normal'' star formation that would be associated with the 
two merging disks independently \citep[see e.g.][]{dimatteo:merger.induced.sb.sims,
cox:massratio.starbursts}. 
Depending on the observational criteria used to identify mergers, of course, 
this definition of star formation ``in mergers'' may be of interest. It is, however, 
a subset of the ``quiescent'' star formation for the most part, and is distinct 
physically (and very distinct in terms of the imprint that it leaves on galaxy 
stellar populations, kinematics, and structural properties) from the short lived, 
compact burst specifically {\em induced} by the merger.

\subsection{AGN and Quasars}
\label{sec:model:mergers.agn}

Given some merger, quasar activity is also excited; 
to lowest order in simulations with AGN-feedback, the peak 
bolometric luminosity of the AGN 
is tightly coupled to the total bulge mass that will be formed from 
a disk-disk merger. The total bulge mass is 
the burst mass (discussed above), plus the violently relaxed 
stellar disk mass, which is simply 
$M_{\rm relaxed} \approx \mu\,M_{\ast}$ 
in simulations and from simple gravitational physics 
considerations \citep[again, see][]{hopkins:disk.survival}. 
At fixed bulge mass, the peak AGN luminosity, corresponding to the 
Eddington limit of the maximum BH mass, is coupled to this 
bulge mass as it must overcome its binding energy in order to 
halt continued growth. In \citet{hopkins:bhfp.obs,hopkins:bhfp.theory}, 
we show how this scales in simulations with bulge mass 
and other properties. We find that it can be conveniently represented 
by the scaling 
\begin{equation}
L_{\rm peak,\,QSO} \approx 4.6\times10^{11}\,\lsun 
\times (1+z)^{0.5}\,
{\Bigl (}\frac{M_{\rm burst}+M_{\rm relaxed}}{10^{10}\,\msun} {\Bigr )}
\ . 
\label{eqn:lpeak.qso}
\end{equation}
The latter scaling simply reflects the fact that the peak/final BH mass scales roughly 
linearly with total bulge mass, in both observations and simulations. 
The $(1+z)^{0.5}$ scaling comes from the simulations discussed in 
\citet{hopkins:bhfp.theory}; it comes from the fact that galaxies at high 
redshift, being both more gas-rich and more compact, require more ``work'' to 
be done by the AGN before it can self-regulate its luminosity/BH mass, 
and so yield higher BH masses at otherwise fixed bulge mass. 
We refer to that paper for more details, but note that the other parameterizations 
of this evolution (discussed therein) yield nearly identical results. 
Likewise, other models for AGN self-regulation at high masses and/or luminosities 
predict a similar maximum BH mass as a function of host galaxy properties 
\citep{silkrees:msigma,murray:momentum.winds,shankar:2009.smbh.demographics.review}, 
and the resulting ``cutoff'' in the AGN luminosities at high masses (owing to 
self-regulation combined with a cutoff in the depth of host galaxy potential wells) 
is similarly predicted in e.g.\ \citet{natarajan:most.massive.bhs}. 
Moreover, such moderate evolution in $M_{\rm BH}/M_{\ast}$ 
is suggested by a number of observations 
\citep[e.g.][and references therein]{peng:magorrian.evolution,
woo06:lowz.msigma.evolution,treu:msigma.evol,
salviander:midz.msigma.evol}. 
Motivated by the simulations and the observed 
BH-host correlations, we assume a constant $0.3\,$dex scatter 
in these relationships.

In what follows, we consider only AGN induced in mergers, but stress that this 
does include non-trivial contributions from minor mergers down 
to e.g.\ mass ratios of $\sim$1:10. 
There is considerable debate regarding whether or not entirely 
non-merger processes such as stellar bars 
\citep[e.g.][and references therein]{
shlosman:bars.within.bars,
jogee:review,younger:minor.mergers}
and/or stochastic encounters with molecular clouds 
\citep{hopkins:seyferts,nayakshin:forced.stochastic.accretion.model}
might drive significant AGN activity.
However, it is generally clear both from observations and 
from simple theoretical considerations that the resulting AGN 
would be important only at low luminosities.
\citet{hopkins:seyfert.limits} compile  
both empirical and theoretical  
estimates of the luminosities below which non-merger 
processes dominate AGN fueling and find
consistently that this is exclusively in the 
traditional Seyfert regime \citep[$L_{\rm bol}\lesssim 10^{12}\,L_{\sun}$; 
see][]{malkan:qso.host.morph,
canalizostockton01:postsb.qso.mergers,
dunlop:qso.hosts,kauffmann:qso.hosts,
floyd:qso.hosts,hutchings:redqso.midz,
zakamska:qso.hosts,
zakamska:mir.seds.type2.qso.transition.at.special.lum,
rigby:qso.hosts,guyon:qso.hosts.ir,
urrutia:qso.hosts}.
This is clear from integral constraints; 
bulges formed in bars (``pseudobulges'') or 
non-merging bulges with bars or central molecular gas concentrations 
dominate only at low galaxy masses, in galaxy types of 
Sb/c and later \citep{kormendy.kennicutt:pseudobulge.review,
fisher:pseudobulge.sf.profile,
allen:bulge-disk,driver:bulge.mfs,
fisher:pseudobulge.ns}. 
Given the observed BH-host correlations, 
this corresponds to BHs with masses $\lesssim10^{7}\,\msun$, or 
maximum luminosities at the Eddington limit of $L_{\rm bol}=3\times10^{11}\,L_{\sun}$. 
But as we will show, AGN are significant in the IR luminosity function 
only at the highest luminosities, $L > 10^{13}\,L_{\sun}$ -- i.e.\ 
BHs with $\sim10^{9}\,\msun$ at Eddington, with accretion rates of 
$>10\,\msun\,{\rm yr^{-1}}$. Since it is unlikely that these 
extreme systems are powered in non-violent events, 
our neglect of non-merger induced AGN 
makes little difference to our predictions. We have, in fact, explicitly 
checked whether including them \citep[according to the model 
luminosity functions predicted in][]{hopkins:seyferts} makes any difference, 
and find it only increases the very low-luminosity contributions of AGN by a factor 
of $\sim2-3$, far less than the $\sim3-4$ orders of magnitude required to substantially 
change our conclusions. Likewise, simply adopting the observed AGN 
bolometric luminosity functions -- including all observed AGN -- 
from \citet{hopkins:bol.qlf} or \citet{shankar:bol.qlf} 
with our estimated template spectra and obscured fractions, we find no
significant difference. 

For a given peak BH mass or peak luminosity, 
we simply require again the distribution of luminosities corresponding to the 
average lightcurve, in order to construct the number density as a function 
of luminosity. 
These lightcurves and the resulting distribution of time spent at different 
AGN luminosities (both bolometric and in various observed bands) 
have been extensively discussed in a series of papers 
\citep{hopkins:lifetimes.methods,hopkins:lifetimes.letter,
hopkins:lifetimes.interp,hopkins:qso.all,
hopkins:faint.slope}. We adopt the Schechter-function 
parameterization therein, 
\begin{align}
\nonumber \frac{{\rm d}t}{{\rm d}\log{L_{\rm bol}}} &\approx 
0.22\,{\rm Gyr}\,
{\Bigl (}  \frac{L_{\rm bol}}{L_{\rm peak}} {\Bigr )}^{\alpha} \,
\exp{{\Bigl \{}-{\Bigl (}  \frac{L_{\rm bol}}{L_{\rm peak}} {\Bigr )}{\Bigr\}}} \\ 
\alpha &\approx -0.44 + 0.21\,\log{(L_{\rm peak}/10^{12}\,\lsun)}
\label{eqn:dt.dl.agn}
\end{align}
Again, this is taken from simulations, but in those papers, it is shown that this 
yields very good agreement with the observed distribution of AGN 
Eddington ratios, host masses and luminosities, and the evolution of the 
AGN luminosity function \citep[most recently, see][]{
volonteri:xray.counts,foreman:binary.quasar.predictions,
bonoli:modeling.qso.clustering.vs.lifetimes.model,marulli:xr.sel.agn.clustering.models, 
hopkins:groups.qso,hopkins:seyfert.limits,hopkins:mdot.dist}.

\subsection{Construction of IR Luminosity Functions}
\label{sec:model:lfs}

Finally, given these predicted SFR and AGN bolometric luminosity 
distributions, we need to convert to the observable quantity, 
namely total infrared luminosity. 
Because we are not attempting to model the 
full SEDs and dust physics of these systems, 
the only quantity that we can robustly predict is the {\em total} infrared luminosity, 
$L_{\rm IR}$, defined as the integrated luminosity from 
$8-1000\,\mu{\rm m}$. 
In the case of SFR distributions, we adopt the simple 
conversion from \citet{kennicutt98}, 
corrected for our adopted \citet{chabrier:imf} IMF, of 
\begin{equation}
L_{\rm IR} = 1.1\times10^{10}\,L_{\sun}\,{\Bigl (}\frac{\dot{M}_{\ast}}{\msun\,{\rm yr}^{-1}}{\Bigr )}\ .
\end{equation}
Note that more sophisticated (e.g.\ luminosity-dependent) 
conversions have been proposed, but since this choice is used to 
calibrate the gas surface density-SFR surface density relation, 
we adopt it for consistency. 
In any case, alternative formulations largely deviate from the above 
only in non-starburst or lower IR-luminosity galaxies, where 
absorption is weaker, but these are not particularly important for our 
comparisons here, and experimenting with those in 
\citet{buat:extinction.vs.LIR} and \citet{jonsson:sunrise.attenuation} yields almost no 
difference at $L_{\rm IR}>10^{11}\,L_{\sun}$. 

An advantage of our semi-empirical model is that our conclusions 
do not depend significantly on the adopted stellar IMF, for typical choices. 
Altering the IMF between e.g.\ \citet{chabrier:imf}, \citet{kroupa:imf}, 
\citet{scalo:imf} 
or e.g.\ \citet{salpeter:imf} generally amounts to systematic changes in
the mass-to-light ratio $M_{\ast}/L$ by up to $0.3\,$dex. However, 
because we begin with observed 
galaxy properties and calculate observed luminosity functions, 
this systematic dependence cancels out. Specifically, 
in adopting the observed \citet{kennicutt98} relation, 
the observable quantity is the luminosity surface density -- this
factor enters in the conversion to a SFR surface density. But then, 
converting the resulting SFR to an observed luminosity, the same factor 
enters, cancelling out. 
Explicitly re-calculating our predictions with other IMF choices confirms this. 
The only residual effects are second order, 
and relatively weak -- for example, changing the implied 
gas exhaustion timescale leads to slightly different dynamics of the gas on 
small scales in mergers. Such details are outside the scope of our 
comparison here, and in any case amount to smaller effects than 
our systematic uncertainties. 
In an {\em a priori} model for star formation, on the other hand, 
the factor would enter fully. 
The IMF will only present a systematic source of uncertainty in our predictions if 
it evolves significantly with redshift or galaxy properties, 
or if it is extremely top-heavy, possibilities we discuss further below.

For AGN, the conversion from bolometric to IR luminosities 
is somewhat more complex -- unobscured (Type 1) 
AGN re-radiate only a fraction $\sim1/40-1/20$ of their bolometric luminosity 
in the FIR, and are thus negligible for the luminosity functions here. 
We adopt the empirically calculated obscured fraction as a function of 
quasar luminosity from \citet{gilli:obscured.fractions}, and assume that 
the obscured bolometric luminosity is re-radiated in the IR; this allows us to convert 
our predicted bolometric QLF to an IR QLF of obscured quasars. 
Technically, not all of the luminosity will be obscured, of course, 
but we find that e.g.\ using the full distribution of column densities as a function 
of quasar luminosity from \citet{ueda03:qlf} to attenuate a template AGN SED yields 
a very similar answer \citep[see also][]{franceschini:faint.xr.qsos}, as does using a mean 
X-ray to IR bolometric correction of obscured AGN \citep{elvis:atlas,
zakamska:multiwavelength.type.2.quasars,polletta:obscured.qsos}. 
The obscured AGN fraction at high luminosities remains uncertain; 
\citet{hasinger:absorption.update} argue that it could be lower by a factor of several 
than the \citet{gilli:obscured.fractions} estimate at the highest luminosities 
(although this is redshift-dependent and at $z\ge1.5$, the two estimates agree well), 
whereas \citet{daddi:2007.high.compton.thick.pops} argue that 
the number of obscured quasars should be a factor of $\sim2-3$ higher. 
These uncertainties are generally comparable to 
(or up to factor $\sim2$ larger than) the uncertainties from the choice of stellar 
mass function, and we discuss some implications below.

\section{Star Formation Rate Distributions and Infrared Luminosity Functions}
\label{sec:lfs}

\subsection{Basic Predictions}
\label{sec:lfs:pred}

\begin{figure*}
    \centering
    \plotside{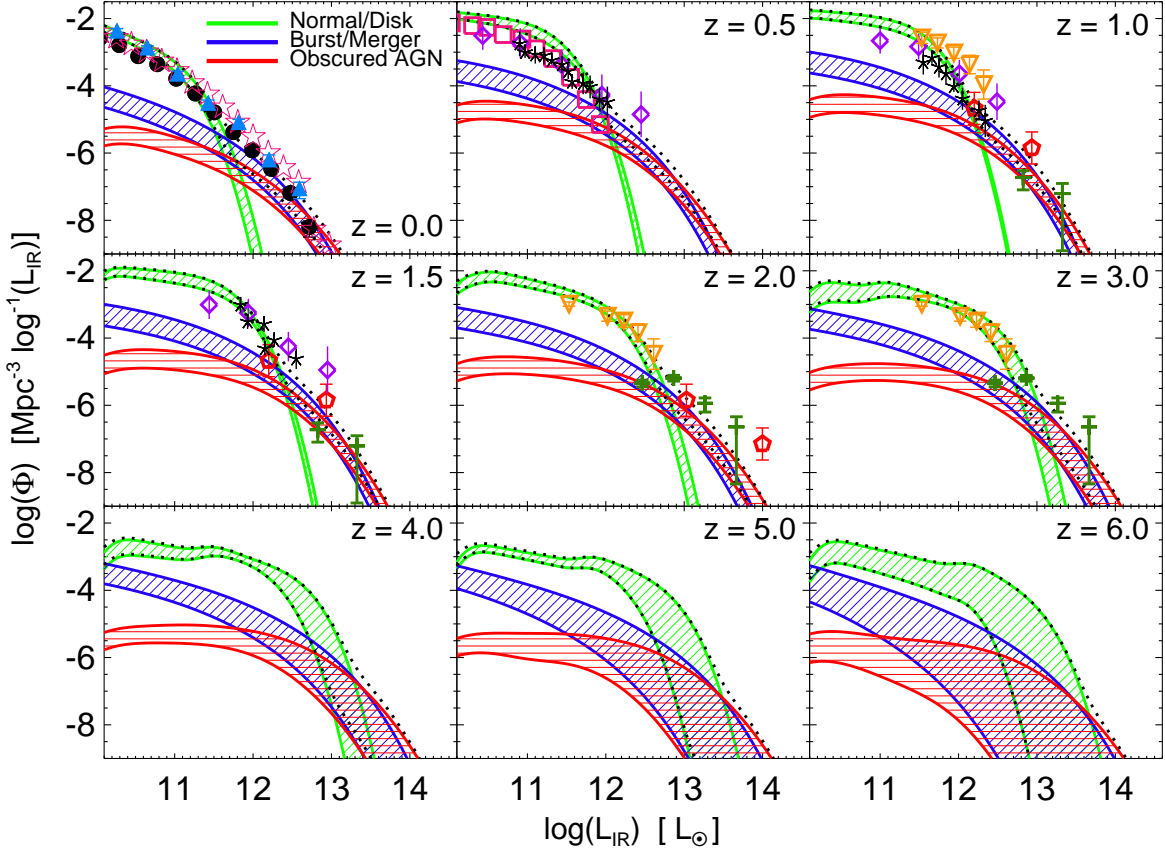}
    \caption{Total ($8-1000\,\mu{\rm m}$) IR luminosity functions as a function of redshift. 
    We show the model contribution from ``normal'' (non-merging) star-forming disks (green), 
    merger-induced starbursts (blue), and obscured AGN (red). 
    The range in the total (summed) LF is shown with dotted black lines. 
    Shaded ranges reflect the 
    uncertainty from different stellar mass function observations used in constructing the model.
    Points show observational estimates from 
    \citet[][magenta stars]{saunders:ir.lfs}, 
    \citet[][blue triangles]{soifer:60m.lfs}, 
    \citet[][black circles]{yun:60m.lfs}, 
    \citet[][magenta squares]{huang:2007.local.ir.lf}, 
    \citet[][violet diamonds]{lefloch:ir.lfs},
    \citet[][orange inverted triangles]{caputi:ir.lfs}.
    \citet[][black $\ast$'s]{magnelli:z1.ir.lfs},
    \citet[][red pentagons]{babbedge:swire.lfs}, 
    \citet[][dark green $+$'s]{chapman:submm.lfs}. 
    \label{fig:ir.lfs}}
\end{figure*}

Figure~\ref{fig:ir.lfs} shows the resulting predicted 
total-IR luminosity functions from $z=0-6$, divided into the contributions 
from ``normal'' (non-merging, quasi-steady-state) star-forming systems, 
merger-induced ``bursts,'' and obscured AGN. 
As discussed in \S~\ref{sec:model}, we have re-calculated our model 
adopting at least two different stellar MF determinations at each redshift; 
the range between the two at each redshift is shown by the shaded 
range, and is representative of the scatter in different observational estimates. 
Unsurprisingly, this uncertainty is substantial at high redshifts. 
We also add (in quadrature) a systematic factor $\sim2$ uncertainty 
in galaxy-galaxy merger rates, representative of the systematic theoretical 
and observational uncertainties as estimated from the compilations 
in \citet{hopkins:merger.rates}. We add a factor $1.5$ additional 
uncertainty in the AGN obscured fractions, again 
representative of systematic observational uncertainties 
\citep[see e.g.][and references therein]{hopkins:lifetimes.obscuration,
hopkins:seyfert.bimodality,shi:silicate.contraints.on.torii,treister:obscured.frac.z.evol,
hasinger:absorption.update}. 
We also note that direct observational constraints used for our 
models of the galaxy stellar mass function are either non-existent or 
extremely uncertain above $z>4$; we extrapolate the fitted LF parameters 
from $z=2-4$ into this redshift range, and so the resulting predictions 
should be treated with the appropriate caution. 

We compare with observations of the IR luminosity functions where available, from 
$z=0-3$. Note that all of these are corrected to a total IR luminosity 
from observations in some band; we adopt the corrections 
compiled in \citet{valiante:ir.lfs.and.numbercounts}, 
but emphasize that some caution, and at least a systematic 
factor $\sim2$ uncertainty in $L_{\rm IR}$, should be considered in 
estimates from most if not all observed wavelengths. 
The agreement between the total predicted LF and the observations is 
generally reasonable, at most redshifts. 
At the highest luminosities and redshifts, 
specifically the sub-millimeter population observed in \citet{chapman:submm.lfs}, 
we appear to under-predict the abundance of bright systems, 
but these observations are very uncertain. 
\citet{austermann:2009.aztec.submm.source.counts}, for example, 
find that the millimeter number counts in these surveys are strongly 
affected by cosmic variance, and may be factors of several larger than the 
cosmic mean. 
We discuss this in \S~\ref{sec:discussion}, but note for now that these systems contribute 
relatively little to the global SFR density at these redshifts. 

At all redshifts, ``quiescent'' galaxies dominate the LF at low luminosities and 
high space densities, reflecting the abundance of star-forming disks 
and relative rarity of mergers and quasars. At higher luminosities, 
eventually merger-induced star formation and AGN activity become dominant, 
as expected in order to explain the most extreme (but short-lived) bursts of 
star formation. However, both types of systems increase in luminosity  
with redshift from $z=0-3$ in similar fashion. 

As discussed in \S~\ref{sec:model:lfs}, the obscured AGN fraction is somewhat 
uncertain. However, Figure~\ref{fig:ir.lfs} shows that even at extreme luminosities, 
the contribution from obscured AGN is comparable to that from merger-induced 
starbursts. Thus, in terms of the total IR luminosity function, even an obscured AGN fraction of 
zero would only lead to factor $\sim2$ changes in the predicted bright-end number 
densities (smaller than the uncertainties owing to the choice of mass function, for 
example). 

To facilitate future comparisons with observations, we present the corresponding 
predictions in Appendix~\ref{sec:appendix:multiwavelength} for the 
IR luminosity function in various specific rest-frame wavelengths. 
However, we stress that these are {\em not} direct predictions of the model -- a 
proper model for the SEDs will depend on full radiative transfer models, applied 
to the simulations as a function of time and galaxy properties 
(these will be presented in future work). 
Here, we simply convert total IR luminosities to wavelength-dependent luminosities 
using the same bolometric corrections used to convert the observations in 
Figure~\ref{fig:ir.lfs}.

\subsection{Fitting Functions to the Predicted LFs}
\label{sec:lfs:fitting}

\begin{figure}
    \centering
    \plotone{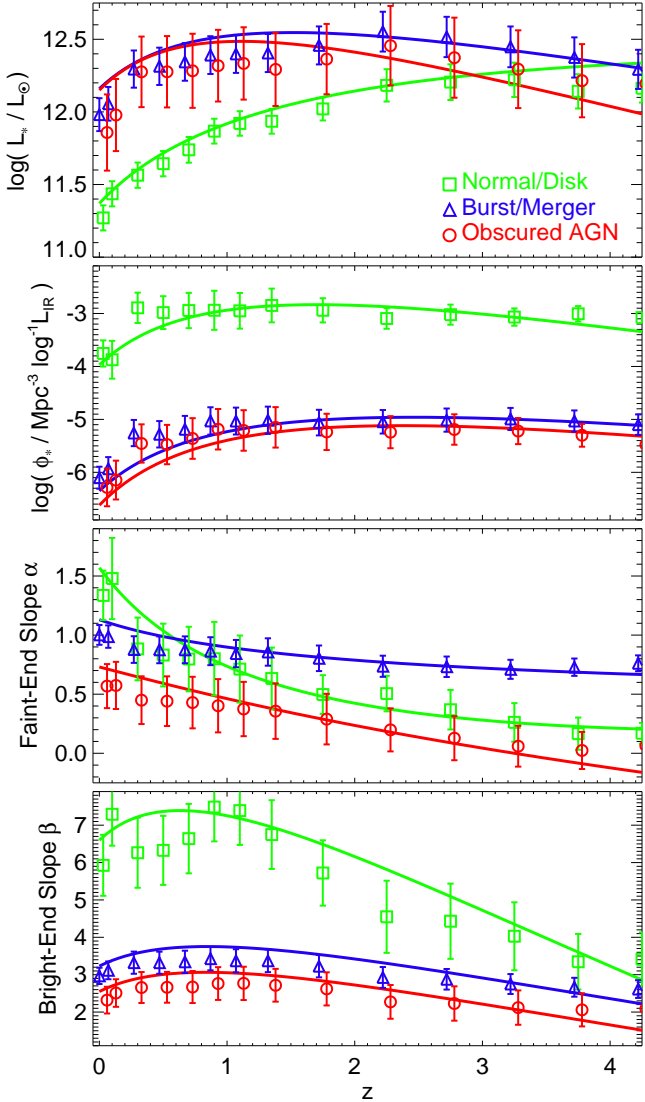}
    \caption{Best-fit parameters for the luminosity functions in 
    Figure~\ref{fig:ir.lfs}, given a double-power law formulation (Equation~\ref{eqn:double.pwr.law}). 
    Points show the best fit (with uncertainty reflecting the allowed range in 
    Figure~\ref{fig:ir.lfs}) at several redshifts, lines the overall maximum likelihood fits
    from Table~\ref{tbl:fits}. Parameter degeneracies are such that the best-fit 
    curves can be systematically slightly offset from the fits at each redshift. 
    {\em Top:} Break luminosity. Rising gas fractions drive the increase; 
    but for mergers, the results asymptote to a maximum owing to the physics of very 
    gas-rich mergers (see text). 
    {\em Second from Top:} Normalization. Modulo a decrease at the lowest 
    redshifts, this is approximately redshift-independent. 
    {\em Second from Bottom:} Faint-end slope. Again, the redshift dependence is weak. 
    The apparent large evolution for star-forming systems is somewhat degenerate with the 
    evolution in the bright-end slope -- a fit where both are held constant is, in fact, 
    acceptable. 
    {\em Bottom:} Bright-end slope. Again, relatively flat with redshift (per note above). 
    Star-forming systems fall off in number density at high luminosities much more steeply 
    than mergers or quasars, reflecting the exponential cutoff in the mass function. 
    \label{fig:lf.fit.results}}
\end{figure}

For the sake of comparison with future observations, we provide fits to the 
model predictions in Table~\ref{tbl:fits}. 
We find that the predicted IR LFs for each type can be reasonably represented 
by a double power-law model, i.e.\ 
\begin{equation}
\Phi \equiv \frac{{\rm d}n}{{\rm d}\,\log{L}} = 
\frac{\phi_{\ast}}{(L/L_{\ast})^{\alpha} + (L/L_{\ast})^{\beta}} 
\label{eqn:double.pwr.law}
\end{equation}
where the parameters $\phi_{\ast}$ (normalization), 
$L_{\ast}$ (break luminosity), $\alpha$ (faint-end slope, 
i.e.\ $\Phi\propto L^{-\alpha}$ for $L\ll L_{\ast}$), 
and $\beta$ (bright-end slope, 
i.e.\ $\Phi\propto L^{-\beta}$ for $L\gg L_{\ast}$) 
depend on redshift, with that dependence conveniently approximated 
as
\begin{align}
\nonumber \log{L_{\ast}} &= L_{0} + L^{\prime}\,\xi + L^{\prime\prime}\,\xi^{2} \\ 
\nonumber \log{\phi_{\ast}} &= \phi_{0} + \phi^{\prime}\,\xi + \phi^{\prime\prime}\,\xi^{2} \\ 
\nonumber \alpha &= \alpha_{0} + \alpha^{\prime}\,\xi + \alpha^{\prime\prime}\,\xi^{2} \\ 
\nonumber \beta &= \beta_{0} + \beta^{\prime}\,\xi + \beta^{\prime\prime}\,\xi^{2} \\ 
\xi &\equiv \log{(1+z)}
\label{eqn:param.z.evol}
\end{align}
(Note that log here and throughout refers to $\log_{10}$.) We perform this 
fit using only our results up to redshift $z=4$, as the HOD constraints used 
to build the model have to be extrapolated at higher redshifts. 
In Table~\ref{tbl:fits}, we quantify the uncertainty in each parameter; 
this reflects the systematic theoretical 
uncertainties shown in Figure~\ref{fig:ir.lfs} (the shaded range), 
with the appropriate covariance between parameters taken into account 
(for this reason, fitting the redshift evolution with free 
parameters up to second-order in $\xi$ leads to relatively 
large uncertainties in the fit results). We also illustrate the best-fit parameters 
as a function of redshift in Figure~\ref{fig:lf.fit.results}. 

The behavior seen in each parameter reflects that discussed above; 
the bright and faint-end slopes, and normalization $\phi_{\ast}$, evolve 
relatively weakly with redshift.\footnote{The apparent ``jump'' in $\phi_{\ast}$ 
at $z\approx0.3$ owes partly to real evolution in the observed input 
mass functions, but mostly to parameter covariance (here between 
$\phi_{\ast}$, $L_{\ast}$, and $\alpha$). Accounting for this 
covariance, the change in $\phi_{\ast}$ from $z=0.2-0.3$ is 
only significant at $1.5-2\,\sigma$, and a smoothly evolving $\phi_{\ast}$ 
provides just as good a fit.}
In fact, we can find reasonable fits within the theoretical 
uncertainties that hold these parameters fixed with 
redshift. But the break luminosity $L_{\ast}$ evolves rapidly, 
as $\propto(1+z)^{2}$ for $z<2$ in all populations, then levels out 
to a maximum at higher redshifts. 
We do see this flattening from $z\sim2-4$, hence the quadratic term here; 
although given our $z<4$ limit ($\log{(1+z)}<0.7$), we are only just 
sensitive to the quadratic terms in $\xi$ (and see no significance fitting 
higher-order terms). 

At all redshifts, $L_{\ast}$ is higher for merger/AGN populations
($>10^{12}\,L_{\sun}$) relative to normal galaxies; but the space density 
$\phi_{\ast}$ is much lower (by a factor of $\sim100-300$). 
The bright-end slope of the normal population is steep, 
reflecting the rapid exponential cutoff in the galaxy mass functions; 
the bright-end slope in the merger/AGN populations is much more shallow, 
$\sim2.5-3$ -- such a slope is, in fact, very similar to the observed 
bright-end slope of the brightest IR populations 
\citep{sanders88:quasars,saunders:ir.lfs,chapman:submm.lfs} 
and to the well-constrained bright-end slope of the 
quasar luminosity functions from redshifts $z\sim0-6$ 
\citep[see e.g.][and references therein]{fan04:qlf,brown06:ir.qlf,richards:dr3.qlf,
hopkins:bol.qlf,shankar:bol.qlf,croom:2009.2slaq.qlf}. 

For comparison, we also consider fitting the LFs to a modified 
Schechter function parameterization, namely
\begin{equation}
\Phi  = \phi_{\ast}\,{\Bigl(}\frac{L}{L_{\ast}}{\Bigr)}^{-\alpha}\,
\exp{{\Bigl\{}-{\Bigl(}\frac{L}{L_{\ast}}{\Bigr)}^{\beta}{\Bigr\}}}\ ,
\label{eqn:modschechter}
\end{equation}
with the same assumed form for the 
evolution in the fit parameters with redshift. 
This is akin to a standard Schechter function except with the 
addition of a bright end ``slope'' term $\beta$, where 
$\beta<1$ allows for a less-steep falloff at high-$L$ than 
would be predicted by a standard Schechter function ($\beta=1$). 
We find that, because the functions shown in Figure~\ref{fig:ir.lfs} 
do not have sharp ``breaks'' characteristic of a double power-law, 
this provides a marginally more accurate representation of the 
LF {\em shape}. 
However, the difference is small, and direct 
interpretation of the parameters in Equation~\ref{eqn:modschechter} 
is complicated by serious fitting degeneracies. 
With this choice of functional form, we find the second-order 
redshift evolution terms make little difference to the fits, 
and so -- given the steep parameter degeneracies involved -- 
do not free the higher-order terms in the fit. 
Interestingly, 
the {\em total} luminosity function obtained by summing 
the contributions from each component is better represented 
with a double power-law, as opposed to the modified 
Schechter function.

\subsection{The Luminosities \&\ Space Densities of Population Transitions}
\label{sec:lfs:transitions}

\begin{figure*}
    \centering
    \plotside{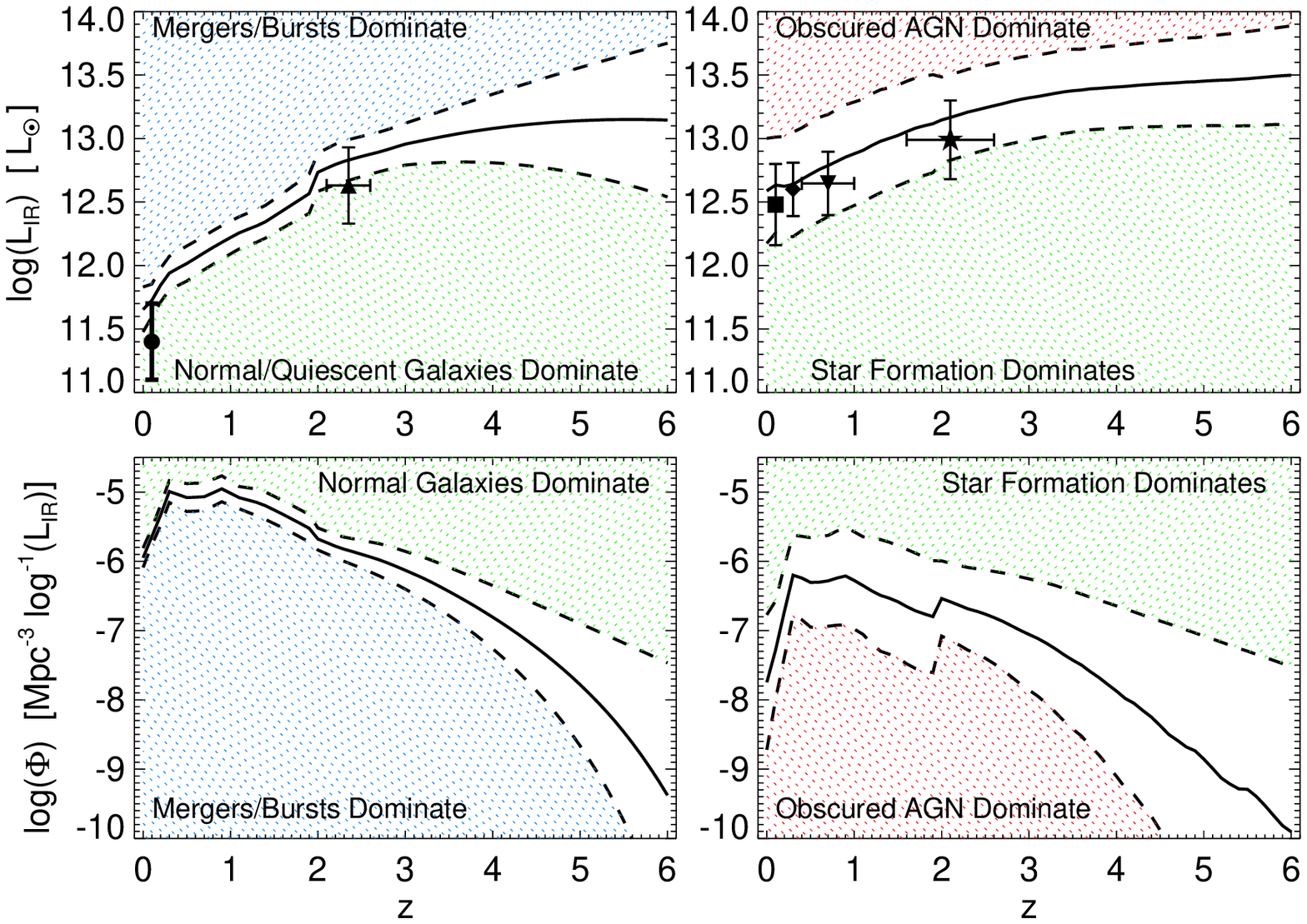}
    \caption{{\em Top Left:} Total IR luminosity threshold above which the predicted 
    IR luminosity functions in Figure~\ref{fig:ir.lfs} transition from being dominated 
    by non-merging (``quiescent'') disks to merger-induced star formation/bursts. 
    We compare observational estimates from morphological 
    studies of bright-IR sources in \citet[][circle]{sanders96:ulirgs.mergers} 
    and \citet[][triangle]{tacconi:smg.mgr.lifetime.to.quiescent}. 
    {\em Bottom Left:} Same, but showing the space density threshold 
    ($\Phi$, in ${\rm Mpc}^{-3}\,\log^{-1}{L_{\rm IR}}$) of the same transition. 
    {\em Top Right:} Luminosity threshold above which the IR luminosity functions 
    transition from being dominated by star formation to being dominated by AGN. 
    Points show the observed estimates from comparison of PAH feature strengths 
    and emission line strength template fitting in 
    \citet[][square]{chary.elbaz:ir.lfs}, 
    \citet[][diamond]{veilleux:ulirg.to.qso.sample.big.mdot.changes,
    veilleux:ir.bright.qso.hosts.merging},
    and \citet[][star]{sajina:pah.qso.vs.sf}, 
    and from comparison of the far IR-radio correlation in \citet{yang:midz.ulirgs}. 
    {\em Bottom Right:} Same, in terms of the space density threshold. 
    As gas fractions increase with redshift, star formation rates increase in all 
    systems. As a result, the threshold for merger-dominance grows from 
    bright LIRGs at $z\sim0$, to bright ULIRGs at $z\sim1-2$, to 
    HyLIRGs at $z>2$. In terms of space density, this transition is relatively constant over 
    this range at $10^{-6}-10^{-5}\,{\rm Mpc^{-3}\,\log^{-1}{L_{\rm IR}}}$ 
    (at higher redshifts, the space densities of all massive systems drop rapidly). 
    At all redshifts, bright HyLIRGs ($L_{\rm IR}\gg 10^{13}\,L_{\sun}$; 
    $\Phi\sim10^{-7}\,{\rm Mpc^{-3}\,\log^{-1}{L_{\rm IR}}}$ at $z\sim0-4$)
    have a non-negligible AGN contribution. 
    \label{fig:transition.lums}}
\end{figure*}

We explicitly quantify the ``transition point'' between the dominance of 
one population or another as a function of redshift in Figure~\ref{fig:transition.lums}. 
Specifically, we define this as the point where the luminosity functions from 
different populations in  
Figure~\ref{fig:ir.lfs} cross. For example, the transition luminosity or space density 
between dominance by normal disks and mergers is given 
by the point in Figure~\ref{fig:ir.lfs} where 
$\phi(L_{\rm IR}\,|\,{\rm normal}) = \phi (L_{\rm IR}\,|\,{\rm burst})$; above this 
$L_{\rm IR}$ (and below the corresponding $\phi(L_{\rm IR})$), 
$\phi(L_{\rm IR}\,|\,{\rm burst}) > \phi(L_{\rm IR}\,|\,{\rm normal})$, 
at lower luminosities and higher space densities the opposite is true. 
Likewise, we can define the 
transition luminosity or space density where obscured AGN become 
numerous than star-formation dominated systems, 
$\phi (L_{\rm IR}\,|\,{\rm AGN})=\phi(L_{\rm IR}\,|\,{\rm normal})+\phi(L_{\rm IR}\,|\,{\rm burst})$. 
The uncertainties in Figure~\ref{fig:ir.lfs} are translated to corresponding 
uncertainties here. 

Our comparisons generally affirm 
the conventional wisdom: at low redshift, mergers dominate the ULIRG and 
much of the LIRG populations, above a luminosity $\sim10^{11.5}\,L_{\sun}$.
Heavily obscured (potentially Compton-thick) 
AGN (in starburst nuclei) become a substantial contributor to IR luminous populations 
in the most extreme $\gtrsim{\rm a\ few\ }\times10^{12}\,L_{\sun}$ systems 
(nearing hyper-LIRG $>10^{13}\,L_{\sun}$ luminosities which are common bolometric 
luminosities for $>10^{8}\,\msun$ BHs near Eddington, but would imply 
potentially unphysical $\gtrsim1000\,\msun\,{\rm yr^{-1}}$ SFRs). 
At higher redshifts, disks are more gas-rich, and thus have characteristically 
larger star formation rates, dominating the IR LFs at higher luminosities. By 
$z\sim1$, most LIRGs are quiescent systems, and by $z\sim2$, only extreme 
systems $\gtrsim{\rm a\ few\ }\times10^{12}\,L_{\sun}$ are predominantly 
mergers/AGN.

This appears to agree well with recent observations. 
First, consider the results of systematic morphological studies of 
IR-bright sources as a function of their luminosities, 
at low redshifts \citep{sanders96:ulirgs.mergers}, which affirms the
conclusion that -- locally -- the brightest LIRGs and essentially 
all ULIRGs are merging systems, while less-luminous systems are not 
(see also references in \S~\ref{sec:intro}). 
At high redshifts, similar studies have now been performed 
\citep[see e.g.][and references therein]{tacconi:smg.mgr.lifetime.to.quiescent}. 
They too find that the brightest sources are almost exclusively mergers, 
but with a transition point (from non-merger 
to merger-dominated) an order-of-magnitude larger in luminosity. 
Other morphological studies at intermediate redshifts $z\sim0.4-1.4$
have reached similar conclusions \citep{bridge:merger.fractions}. 

Other studies have attempted to separate the contributions of star 
formation and (obscured) AGN. At low redshifts, we find similar 
results from observational comparison of emission line strengths 
\citep{sanders96:ulirgs.mergers,kewley:agn.host.sf}, 
observations of the strength of observed PAH features 
\citep{lutz:pah.qso.vs.sf.local,veilleux:ulirg.to.qso.sample.big.mdot.changes}, 
full SED template fitting 
\citep{farrah:qso.vs.sf.sed.fitting}, or indirect comparison with Type 2 AGN luminosity 
functions \citep{chary.elbaz:ir.lfs}. In each case, 
these studies find that local ``normal'' ULIRGs are star-formation dominated, 
but extremely rare systems approaching Hyper-LIRG luminosities tend to be 
AGN-dominated. At high redshifts, there have recently 
been attempts to apply similar methodologies, especially comparison of 
PAH strengths, and we show such an estimate from \citet{sajina:pah.qso.vs.sf}, 
who find a similar transition from star formation to AGN dominance as at low redshift, 
at a factor of several higher luminosity. 
\citet{yang:midz.ulirgs} measure dust temperature distributions and positions on 
the far IR-radio correlation for a sample of ULIRGs over the redshift 
range $z=0.3-1$; they find that below $L_{\rm IR}=10^{12.4}\,L_{\sun}$ 
($L_{\rm FIR}=10^{12.25}\,L_{\sun}$), 
the systems appear star-formation dominated, while above 
$L_{\rm IR}=10^{12.9}\,L_{\sun}$, the IR luminosities are 
dominated by AGN. 
At $z=2$, the same constraints 
support the conclusion from \citet{sajina:pah.qso.vs.sf}; 
\citet{younger:mm.obs.z2.ulirgs} show that samples of $\sim2-8\times10^{12}\,L_{\sun}$ 
ULIRGs at $z=2$ follow the local far IR-radio correlation, indicating 
they are starburst dominated, but \citet{bussmann:2009.dog.luminosities,
bussmann:2009.dog.agn.morphologies} find that by 
luminosities of $\sim2\times10^{13}\,L_{\sun}$, 
IR samples are dominated by warm dust sources 
more likely to be (post-merger) AGN. Of course, changing the assumed 
number of obscured AGN, as a function of luminosity or redshift, will 
correspondingly shift the predicted transition point; the agreement seen 
here suggests that the correct number is probably not very different 
from that adopted here. 
For further details, we refer to the above as well as
\citet{chapman:submm.lfs}, \citet{dey:2008.dog.population}, 
and \citet{casey:highz.ulirg.pops}.

The transition point between non-merger and 
merger dominance of the luminosity function shifts to larger luminosities 
at high redshifts, even though gas-rich merger rates increase rapidly. 
The evolution in gas fractions, which drives up both disk star formation rates 
and merger-induced bursts similarly, is the dominant effect; 
the evolution in merger rates is also not {\em so} rapid as to  
dominate the population at redshifts $z\lesssim2$. 
Moreover, as noted in \S~\ref{sec:model}, although disk star formation 
rates (at otherwise fixed properties) 
increase monotonically with their gas fractions and hence gas surface densities, 
merger-induced bursts and quasar episodes can decline in efficiency in 
extremely gas-rich systems, because the gravitational torques that allow 
for such bursts depend on a sizable dissipationless (stellar) disk 
component \citep{hopkins:disk.survival}. As a result, merger-induced bursts 
do not grow in importance as rapidly as might naively be expected from 
analysis of e.g.\ halo-halo merger rates.

\subsection{Corresponding SFRs and Bolometric Luminosities}
\label{sec:lfs:sfr}

\begin{figure}
    \centering
    \plotone{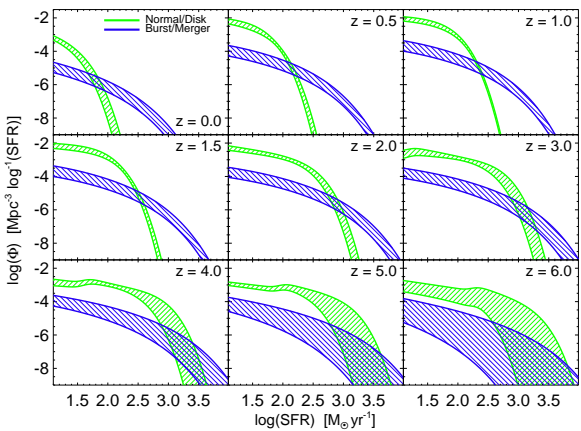}
    \caption{As Figure~\ref{fig:ir.lfs}, but showing the distribution of 
    star formation rates at each redshift (hence no AGN contribution). 
    \label{fig:sfr.lfs}}
\end{figure}
\begin{figure}
    \centering
    \plotone{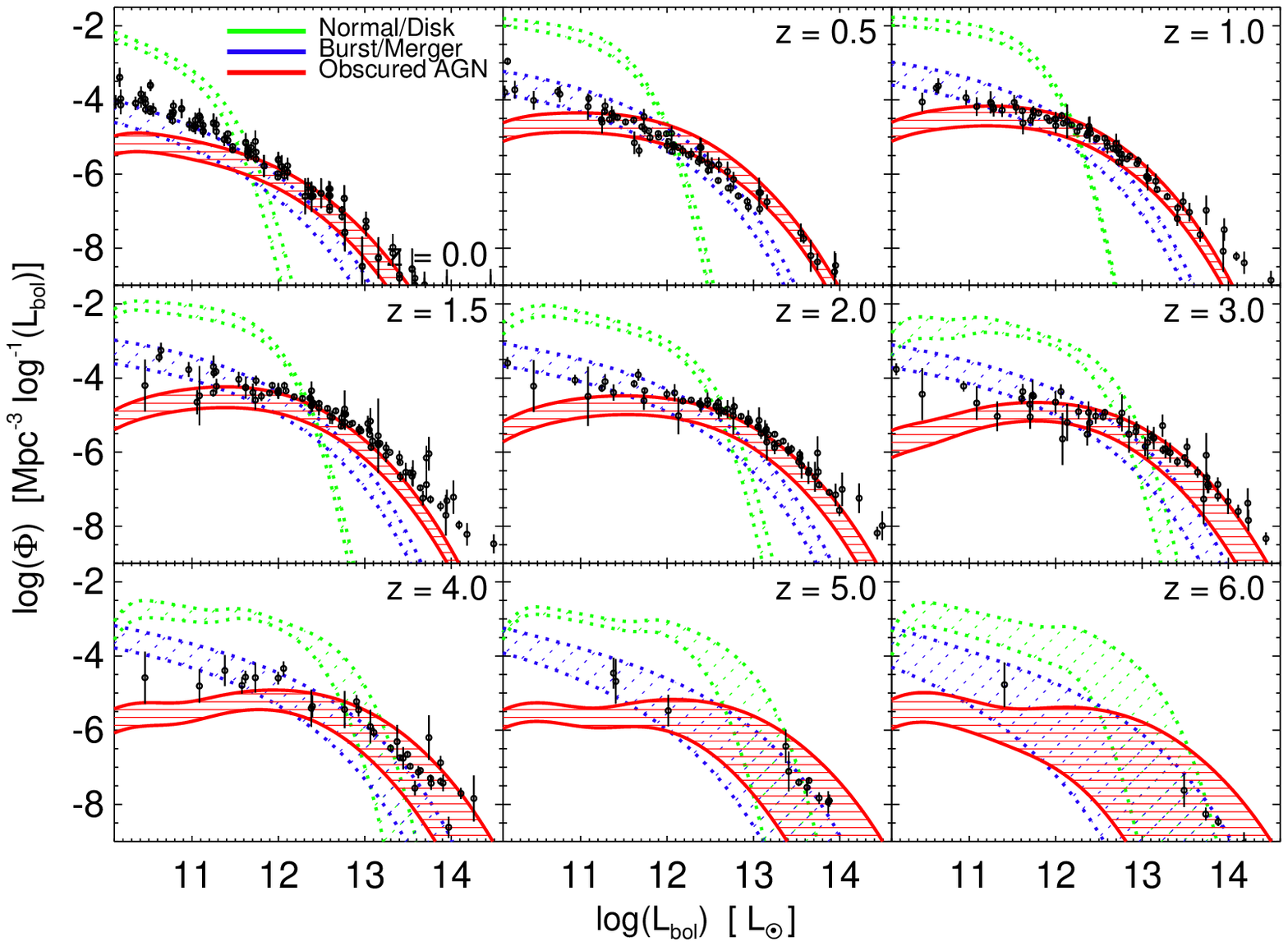}
    \caption{As Figure~\ref{fig:ir.lfs}, but in terms of bolometric luminosity 
    (similar to the IR for star-forming systems, but significantly larger for AGN, 
    which are highlighted here). 
    Black points show the compilation of observational data used to derive 
    bolometric AGN luminosity functions in \citet{hopkins:bol.qlf}; these should 
    be compared to the predicted AGN LF. 
    \label{fig:bol.lfs}}
\end{figure}

In Figure~\ref{fig:sfr.lfs} we reproduce our model from Figure~\ref{fig:ir.lfs}, 
but show instead the distribution of star formation rates. Of course, AGN are 
not present here, since although there will be star formation in their hosts, 
the AGN IR luminosity itself is not from star formation. 

Figure~\ref{fig:bol.lfs} shows the corresponding bolometric luminosity 
functions. For the star-forming systems, this is essentially identical to the 
total-IR luminosity functions, as the total IR emission dominates the 
bolometric luminosity in at least the luminous, IR-bright end of the 
distribution (of interest here). 
However, for the AGN, the bolometric emission 
is considerably larger than the IR emission. We therefore highlight the 
AGN predictions. We compare these to the large compilation of observations 
used to derive bolometric quasar luminosity functions in \citet{hopkins:bol.qlf} 
(see references therein). Similar results have been obtained in other compilations 
\citep{shankar:bol.qlf}, or from hard X-ray luminosity functions with 
appropriate bolometric corrections \citep[see e.g.][]{aird:hx.qlf,yencho:hx.qlf}. 
The observationally estimated bolometric QLF agrees reasonably well with 
our predictions. The under-prediction of very low-luminosity AGN 
owes to our neglect of non-merger induced AGN (\S~\ref{sec:model:mergers.agn}); 
it is clear here that these have a negligible impact on our conclusions. 
The model may also somewhat under-predict the number density of the most 
luminous systems ($L_{\rm bol}>10^{14}\,L_{\sun}$); this is discussed 
in detail in \citet{hopkins:groups.qso}, but is sensitive to the assumed scatter 
in bolometric corrections, and to the existence of 
even a small lensed or beamed QSO population.

\subsection{The Effects of Different Model Assumptions}
\label{sec:lfs:tests}

We briefly outline the effects of several important components in the models 
adopted. For further details, see Appendix~\ref{sec:appendix:assumptions}, 
where we reproduce our model from Figure~\ref{fig:ir.lfs} explicitly, with different 
changes (discussed here) to the model. 

If we do not allow galaxy gas fractions to evolve with redshift (i.e.\ adopt the $z=0$ 
value at all redshifts), this leads to a substantial under-prediction of the luminosities of 
``quiescent'' galaxies at $z\ge 1$. In 
short, the existence of apparently ``normal'' galaxies at high redshifts, 
with ULIRG-level luminosities, requires very high gas surface densities 
(relative to those at $z=0$) if the \citet{kennicutt98} relation is to hold in some form. 

If we do not allow for disk sizes to be more compact at high redshift, this yields lower 
surface densities, and hence somewhat lower SFRs, but the effect 
is relatively minor. Because the observed size evolution (of star-forming galaxies) is weak 
($R_{e}(M_{\ast})\propto(1+z)^{-(0-0.6)}$) and the size evolution (at otherwise fixed properties) 
only enters into the SFR at sub-linear order (for $\dot{\Sigma}_{\ast}\propto\Sigma_{\rm gas}^{1.4}$, 
this yields $\dot{M}_{\ast}\propto R^{-0.8}$ at otherwise fixed properties), 
the total difference is relatively small (factor $\sim 2$) in luminosity, comparable to many of 
the other uncertainties involved. 

If we do not allow for 
scatter in any quantities (e.g.\ disk sizes, gas fractions, burst masses, quasar 
bolometric corrections, and SFRs at otherwise fixed properties) -- i.e.\ force all 
values to exactly trace the medians given in \S~\ref{sec:model} -- this has the 
expected effect, that the rare, high-$L$ population is significantly suppressed. These
objects depend on the existence of some systems with relatively high 
gas fractions at high masses and high $L_{\rm IR}$ relative to many of their 
other properties. 

We can also re-construct our model predictions, but 
adopt a more steep power-law index for the \citet{kennicutt98} relation, for example 
$\dot{\Sigma}_{\ast}\propto\Sigma_{\rm gas}^{1.6}$, 
as suggested by some recent observations of high-redshift systems 
\citep{bouche:z2.kennicutt}. In order to avoid over-producing local star formation 
rates (and indeed the LFs at all luminosities and redshifts), it is necessary to 
correspondingly re-normalize the relation: we do so such that 
a Milky-Way like disk, with effective gas surface density 
$\approx 3\times10^{8}\,\msun\,{\rm kpc^{-2}}$ ($10\%$ gas fraction)
has the same SFR as that expected from the relation fit by \citet{kennicutt98}. 
This amounts to a factor $\approx 3.1$ lower normalization 
in Equation~\ref{eqn:kennicutt}, with the steeper 
$n_{K}$. Considering simulations with such a steeper index, the resulting 
burst properties are qualitatively similar, but the burst timescale in 
Equation~\ref{eqn:tburst} is shorter by a factor $\approx2$,  
$t_{\rm burst}\approx0.4\times10^{8}\,$yr. 
The AGN properties are relatively unchanged. 
Together, the results from this revised 
model are similar to our default model -- however, the steeper 
index leads to more star formation in the very high gas density systems 
at high redshift. This actually somewhat improves the agreement with the 
observed number densities of the most luminous systems; however, 
the difference is ultimately within the range of our other uncertainties, 
in particular the number density of the most massive galaxies. 

Again, to facilitate future comparisons, we provide in 
Table~\ref{tbl:fits} fits to the same double-power law functional 
form for the predicted luminosity functions in both the case of 
no gas fraction evolution, and the case of a steeper 
Kennicutt-Schmidt index.

\subsection{The Luminosity/SFR Density: Contribution of Mergers and AGN}
\label{sec:lfs:density}

\subsubsection{Luminosity Densities: Predictions}
\label{sec:lfs:density:pred}

\begin{figure}
    \centering
    \plotone{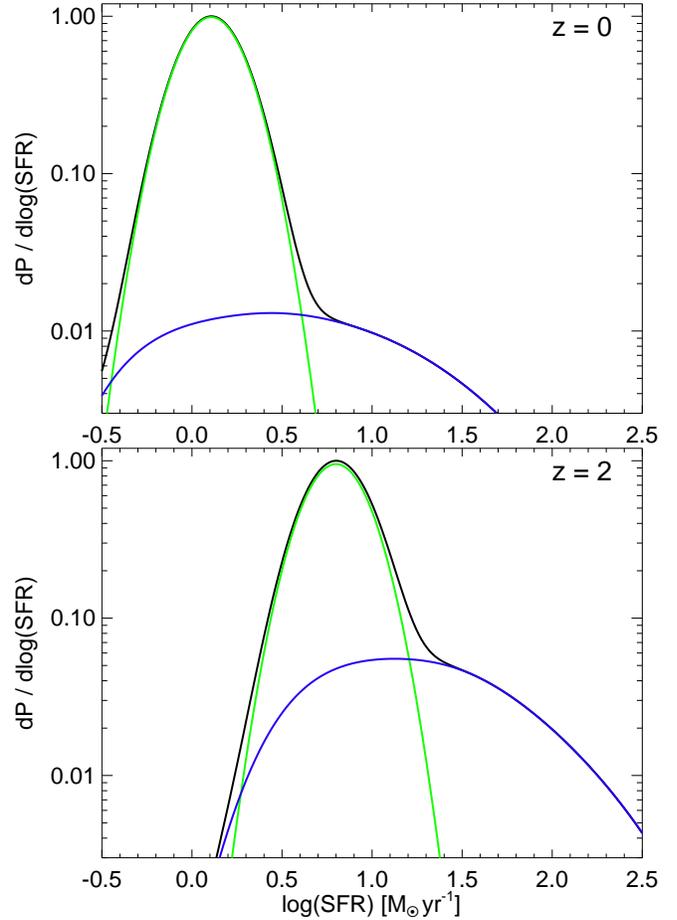}
    \caption{Distribution of SFR in galaxies of fixed mass ($M_{\ast}=10^{11}\,\msun$) at 
    $z=0$ and $z=2$. We show the total (black), contribution from ``normal'' systems 
    (green), and contribution from merger-induced bursts (blue). The ``normal'' systems here 
    have scatter reflecting that in their radii and gas fractions (assumed lognormal). 
    Mergers, even at $z=2$, do not dominate the population or scatter in SFR($M_{\ast}$), 
    except in the extreme wings, consistent with observations indicating small deviations about 
    the median $M_{\ast}$-SFR trend in normal galaxies \citep{noeske:sfh}. Observations 
    probing between $\sim2-3\,\sigma$ in the wings are needed to see the merger ``tail.''
    \label{fig:sfr.scatter}}
\end{figure}

In Figure~\ref{fig:sfr.scatter}, we examine the distribution of 
SFRs at fixed galaxy stellar mass (for an $\sim L_{\ast}$, 
$M_{\ast}=10^{11}\,\msun$ system), for the standard model used in 
constructing Figure~\ref{fig:ir.lfs}, at $z=0$ and $z=2$. 
We separately show the distribution of SFRs from the ``quiescent'' 
(non-merger) systems at that mass, and from the merger-induced bursts. 
The scatter in non-merger systems comes from the distribution of 
gas fractions and effective radii, at a given stellar mass and redshift. 
Obviously, SFRs are systematically higher at $z=2$, 
and the merger contribution is relatively larger, 
as merger fractions observed have increased from $\sim1\%$ at 
$z=0$ to $\sim10\%$ at $z=2$ \citep[see e.g.][and references therein]{
bundy:merger.fraction.new,conselice:mgr.pairs.updated,
kartaltepe:pair.fractions,lin:mergers.by.type,bluck:highz.merger.fraction,
hopkins:merger.rates,jogee:merger.density.08,
bridge:merger.fraction.new.prep}.
However, at both redshifts, the merger contribution is 
relatively small, and although it dominates the tail at very high SFR at fixed 
mass, it constitutes much less than the $>30\%$ of the population 
needed for it to bias the $\sim1\,\sigma$ scatter in SFR$(M_{\ast})$. 
Various observations have shown that there is a tight sequence of 
SFR with galaxy mass in star-forming systems \citep[e.g.][]{noeske:sfh}, 
with small scatter $\lesssim0.3$\,dex, similar to that predicted here. 
This presents a constraint on the role of merger-induced bursts in affecting 
SFRs, but one easily satisfied here -- far from affecting the scatter 
at the $1\,\sigma$ level, one has to observe the scatter at a level 
between $2-3\,\sigma$ in the high-SFR ``wings'' of the distribution at 
fixed stellar mass before the merger-induced tail would be evident.

\begin{figure*}
    \centering
    \plotside{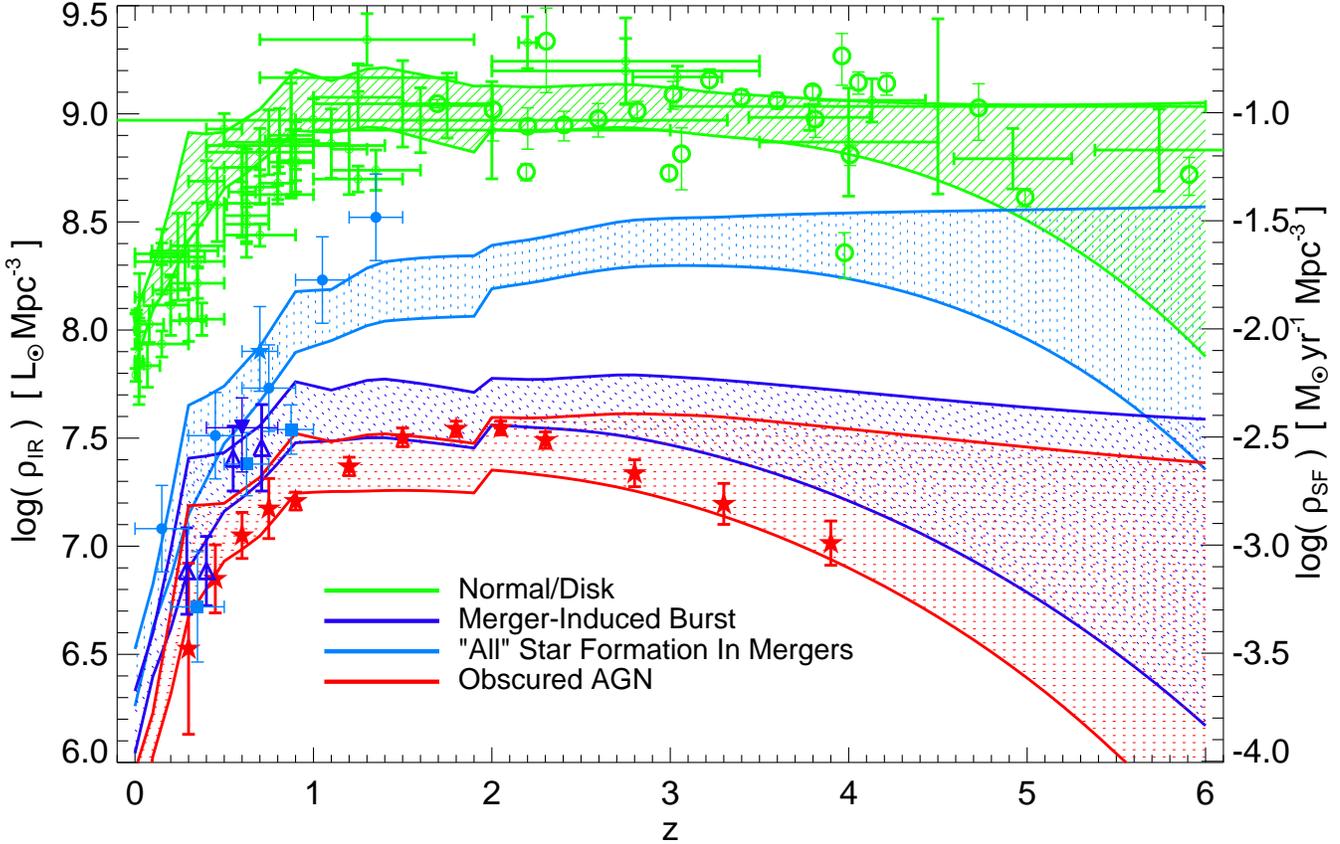}
    \caption{Total IR luminosity density (and corresponding SFR density) as a function of redshift. 
    We show the model prediction for the contribution from ``normal'' (non-merger-induced) 
    disk star formation (green), and that for merger induced bursts (dark blue), and 
    AGN (red), as in Figure~\ref{fig:ir.lfs}. We also show the total IR luminosity density associated 
    with ``ongoing'' interactions, which might be identified via morphology or pair-selected 
    samples for a duration $\sim1\,$Gyr (light blue). Note that most of the star formation in these systems 
    is the continuation of their ``quiescent'' star formation -- the specifically merger-induced 
    starburst lasts $\sim10^{8}\,$yr. 
    We compare the observational compilation from \citet{hopkinsbeacom:sfh} 
    for the total IR luminosity density/SFR density (green diamonds) 
    and at high redshifts the SFR density inferred from Lyman-$\alpha$ forest 
    measurements in \citet[][green circles]{faucher:ion.background.evol}. 
    We also compare 
    the results from the multi-wavelength ``bolometric'' AGN luminosity function 
    compilation from \citet{hopkins:bol.qlf} for the IR luminosity density in AGN (red). 
    Observational estimates of the SFR density specifically {\em induced} 
    by mergers (e.g.\ subtracting some ``baseline'' SFR estimate 
    from a control sample of non-merging systems for identified mergers) 
    are shown (dark blue) from 
    \citet[][triangles]{jogee:merger.density.08} and 
    \citet[][inverted triangle]{robaina:2009.sf.fraction.from.mergers}; 
    estimates of the total SFR density in ongoing/identifiable 
    (e.g.\ pair or morphologically selected) mergers 
    are shown (light blue) from 
    \citet[][circles]{menanteau:morphology.vs.sfr},
    \citet[][squares]{brinchmann:1998.morph.contrib.to.sfr}, 
    and \citet[][star]{bell:morphology.vs.sfr}.
    At all redshifts, AGN represent a small contribution to the total 
    (FIR-dominated) IR luminosity density. Merger-{\em induced} star formation is a 
    similarly small $\sim 5\%$ of the IR luminosity density; the {\em total} luminosity 
    density associated with mergers is somewhat larger, but still small, rising 
    from $\sim5-10\%$ at $z<1$ to $\sim10-20\%$ at $z\sim2-4$. 
    \label{fig:lumden}}
\end{figure*}

Figure~\ref{fig:lumden} combines the LFs predicted in Figure~\ref{fig:ir.lfs} 
to show the total infrared luminosity density, and corresponding total 
SFR density, of the Universe as a function of redshift. 
Approximate fits to these predictions can be obtained 
by simply integrating the fitted LFs in Table~\ref{tbl:fits}. 
The luminosity density in quiescent systems dominates the global total at all redshifts. 
Merger-induced bursts contribute a relatively small fraction to the 
global SFR density; rising from $\sim1-5\%$ at low redshifts $z\sim0$ 
to a roughly constant $\sim4-10\%$ at $z>1$.

The contribution from obscured AGN is at most comparable to that 
from merger-induced bursts, and in general a factor of $\sim2-3$ lower 
(of course, the conversion to SFR density is not valid for AGN, 
as the IR emission is powered by accretion; they should be compared to the 
total IR luminosity density only). 
It is unlikely that obscured AGN contribute 
more than $\sim5\%$ to the global IR luminosity density, even assuming a 
generous near-isotropically obscured fraction of $\sim1/2$ 
at high luminosities \citep[large given the observational constraints 
from e.g.][]{gilli:obscured.fractions,tajer:xr.optical.obscured.agn,
daddi:2007.high.compton.thick.pops,hickox:bootes.obscured.agn,
caccianiga:true.obscured.agn.fracs.xbs.dilution,treister:obsc.frac.from.midir.excess,
menendezcelmestre:extended.submm.galaxies,treister:compton.thick.fractions,
malizia:integral.obscured.agn.column.dist,trichas:2009.sb.agn.lir.vs.lx}.
In fact, allowing the entire {\em bolometric} AGN luminosity density 
estimated in \citet{hopkins:bol.qlf} and \citet{shankar:bol.qlf} to be re-radiated 
in the IR increases the contribution of AGN by only a factor $\sim3$. 
Even under conservative assumptions, then, the contribution from obscured 
AGN is much less than the other current statistical and systematic 
errors in the estimation of the global SFR density, and infrared-derived 
SFR densities are not likely to be significantly contaminated by AGN. 
This question has been studied via other means, as well -- for example, in 
X-ray background synthesis models -- with similar conclusions 
\citep[see e.g.][]{treister:obscured.frac.z.evol,treister:obsc.frac.from.midir.excess}. 

These are global statements -- at a given (high) luminosity, the contribution of merging 
systems and/or AGN may be much higher. Moreover, at any specific 
frequency, the results here could be quite different -- for example, 
in near and mid-IR wavelengths, AGN might be relatively much more 
luminous than cold dust emission from galaxy-wide starbursts 
(and even un-obscured AGN will contribute significantly at these 
wavelengths), thus the AGN luminosity density in such a rest-frame 
band might compete with or dominate the luminosity density 
from star formation \citep[see e.g.][]{blain:ir.lf.synthesis.model}. 

As discussed in \S~\ref{sec:model}, there is an important difference between 
the total SFR density specifically {\em induced} via merger-driven 
galaxy starbursts (i.e.\ gas losing angular momentum owing to gravitational 
processes in the merger, falling to the galaxy center, and driving a 
short-lived starburst over $\sim10^{8}\,$yr timescales), 
and the total SFR density that might be identified observationally 
as ``in ongoing mergers.'' The latter includes all star formation 
in systems that would be identified as merging (usually 
specifically limited to ``major'' mergers), via either some morphological 
or pair-separation based selection criteria. 
The duration of these phases (and hence the total SFR density associated 
with mergers in such a manner) depends on the exact 
selection criteria, but calibration of observational methodologies with 
numerical simulations suggests it is $\sim1\,$Gyr \citep[see e.g.][]{lotz:merger.selection}. 
During this time, except for the much shorter duration of the burst itself, 
the SFRs will (to lowest order, at least) reflect the 
``quiescent'' or ``normal'' SFR of the disks, appropriate for their gas content and 
structural properties \citep[see e.g.][]{dimatteo:merger.induced.sb.sims,
cox:massratio.starbursts,cox:winds.prep}. 
In order to compare with these observations, we calculate the analogous 
SFR density in ongoing mergers with the following simple method: 
given the total rate of major mergers at each redshift, we simply assume a 
$1\,$Gyr observable lifetime for each such (major, $\mu>1/3$) merger, 
and during this time assume it has a constant SFR equal to the rate of 
the quiescent systems with the same properties. We add the SFR density 
calculated in this fashion to that from the bursts themselves, and obtain 
an estimate of the ``total'' SFR that might be associated with e.g.\ disturbed or 
paired systems. 

This is much larger than the burst SFR density, especially at high 
redshifts. At high redshifts, merger rates are high, so a long observable 
duty cycle $\sim1\,$Gyr means that a large fraction of systems will appear 
perturbed (i.e.\ the ``merger fraction'' will become large), and so a large 
fraction of star formation will appear in mergers. Here, we estimate 
this to rise from a few percent at $z=0$ to $\sim20\%$ at $z\sim2$ and 
as high as $\sim20-50\%$ at $z\sim3-6$. 
This is similar to the conclusions from the analysis in \citet{hopkins:merger.lfs}, 
using a different methodology but similarly attempting to calculate the 
total SFR in ``ongoing'' mergers. 
Of course, this should be larger than the burst SFR density at all times; 
but the primary reason the difference becomes so large at high redshift 
is that typical gas fractions are very large. As discussed in \S~\ref{sec:model}, 
and shown in detail in simulations in \citet{hopkins:disk.survival,hopkins:disk.survival.cosmo}, 
large gas fractions lead to less efficient angular momentum loss and so 
(relatively) less efficient bursts in mergers, on average. However, local star formation 
in disks is not affected by this; star formation rates in the ``quiescent'' or extended 
disk mode continue to rise super-linearly with $f_{\rm gas}$ according to the 
\citet{kennicutt98} law.

\subsubsection{Luminosity Densities: Comparison with Observations}
\label{sec:lfs:density:obs}

We compare these predictions to a number of observational 
constraints. First, for the total luminosity/SFR density, we 
show the compilation of observations presented in 
\citet{hopkins:sfh,hopkinsbeacom:sfh}. These come, for the SFR 
density, from a variety of different observations at various wavelengths; 
they are shown here in terms of the estimated total SFR density with 
the IR luminosity density following from the standard conversion adopted here. 
Complementary constraints at higher redshift can be inferred 
from observations of the Lyman-$\alpha$ forest 
in quasar spectra and ionizing background, compiled 
in \citet{faucher:ion.background.evol}. 
At all redshifts, the prediction is 
within the scatter of these observations; at low redshifts $z\lesssim1$, 
the median predicted is somewhat higher than the median of the observed points 
but the difference is small in absolute terms, $\sim0.2\,$dex -- well within 
the systematic uncertainties of both theory and observations. 

Next, we consider the luminosity density in obscured AGN. 
\citet{hopkins:bol.qlf} present bolometric quasar luminosity functions, 
compiled from observations at a wide range of different wavelengths, 
together with observationally inferred column density distributions and 
template spectra. Adopting the fits therein,\footnote{
A code for generating the observed quasar luminosity functions in 
various bands, based on these observations, is provided 
at \qlfcalcurl.} assuming that 
the obscured luminosity is re-radiated in the FIR, we construct the 
corresponding QSO IR luminosity density. 
This agrees well with our theoretical estimate. 
Similar constraints are obtained from the complimentary QLF 
compilations presented in \citet{shankar:bol.qlf}, 
and from synthesis models of the IR backgrounds 
\citep[e.g.][]{blain:ir.lf.synthesis.model}. 

We also compare observational estimates of the 
luminosity/SFR density in mergers. 
First, several authors have attempted to estimate 
the total amount of star formation in observationally 
identified ongoing mergers or recent (morphologically disturbed) 
merger remnants. We compare observations 
compiled from \citet{brinchmann:1998.morph.contrib.to.sfr,
menanteau:morphology.vs.sfr,
bell:morphology.vs.sfr}, 
who estimate this quantity in morphologically-selected samples 
at $z\sim0-1.5$.\footnote{Note that most of these authors 
actually measure the {\em fraction} of the SFR density in 
or induced by mergers, not the absolute value. We convert this 
to an absolute density by rescaling with the observed 
total SFR density at the same redshift from the best-fit observed 
trend presented 
in \citet{hopkinsbeacom:sfh}. Since this agrees well with our predicted 
total SFR density, it makes little difference if we use that instead.}
Second, more recently, attempts have been made to specifically 
isolate the merger-{\em induced} star formation rate density. 
Typically, in these cases, the SFR density of some merger 
sample (identified in a similar manner) is considered, but 
only after subtracting away/removing the contribution from 
the expected ``normal'' mode star formation. In general, this is accomplished via 
comparison to some control sample of star forming galaxies with similar 
stellar masses and redshifts. \citet{robaina:2009.sf.fraction.from.mergers} attempt this 
from a pair-selected sample at $z\sim0.4-0.8$; 
\citet{jogee:merger.density.08} consider a similar estimate from 
morphologically selected samples at $z\sim0.4-1$. 
Clearly, the two estimates (total SFR in ``ongoing'' mergers versus the 
SFR density {\em enhancement} from mergers) should be 
compared to the appropriate respective theoretical predictions, 
as discussed above. In both cases we see good agreement. 
At $z\sim1$, the observations may in fact indicate the predicted 
growing difference between ``all'' star formation in mergers 
and the star formation specifically induced by mergers.

\subsubsection{Contributions from LIRGs and ULIRGs}
\label{sec:lfs:density:contribs}

\begin{figure}
    \centering
    \plotone{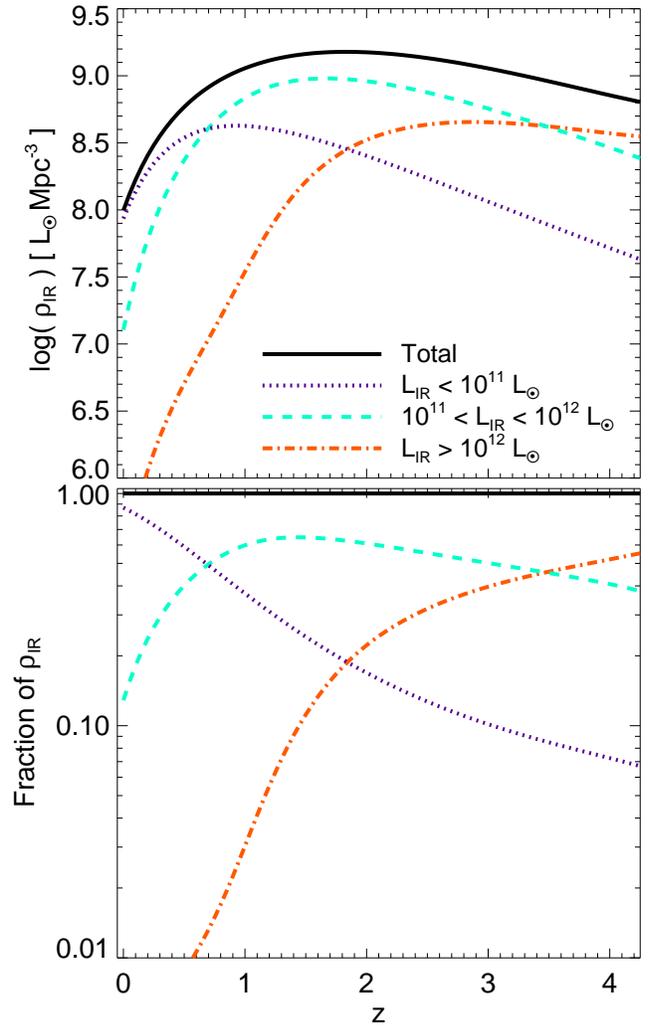}
    \caption{Contribution to the IR luminosity density from 
    galaxies in different luminosity intervals. 
    {\em Top:} Luminosity density, as Figure~\ref{fig:lumden}. 
    We compare the total, and contribution from 
    sub-LIRG ($L_{\rm IR}<10^{11}\,L_{\sun}$), 
    LIRG ($L_{\rm IR}=10^{11}-10^{12}\,L_{\sun}$), 
    and ULIRG ($L_{\rm IR}>10^{12}\,L_{\sun}$) systems. 
    HyLIRGs ($L_{\rm IR}>10^{12}\,L_{\sun}$) are negligible ($\lesssim1-5\%$ contribution)
    in this total at all redshifts. 
    For clarity, we show the results for the fits in Table~\ref{tbl:fits}; 
    each curve has an approximate $0.15$\,dex uncertainty. 
    {\em Bottom:} Same, as a fraction of the total $\rho_{\rm IR}$. 
    LIRGs rise to dominance by $z\sim1$. ULIRGs are comparable and 
    then dominant in output at $z\sim2$ and $z\sim3$, respectively. 
    We stress that, as in Figure~\ref{fig:transition.lums}, 
    these luminosity cuts do {\em not} necessarily correspond 
    to physically different classes of systems. 
    \label{fig:lumden.by.L}}
\end{figure}

In Figure~\ref{fig:lumden.by.L}, we illustrate the contributions 
to the luminosity density from galaxies in various luminosity 
intervals -- specifically, non-LIRG, LIRG, and ULIRG systems. 
For clarity we show 
just the results from our best-fit luminosity functions in 
Table~\ref{tbl:fits}; the full allowed range scatters about these 
curves by $\sim0.15\,$dex. 
We find the well-known result from a number of observational studies 
\citep[see e.g.][]{lefloch:ir.lfs}: higher-luminosity systems progressively 
dominate more of the IR luminosity density at higher redshifts. 
At $z=0$, the luminosity density is dominated by 
relatively low-luminosity $L_{\rm IR}\sim10^{10}\,L_{\sun}$ systems.
The contribution from 
LIRGs ($L_{\rm IR}>10^{11}\,L_{\sun}$ systems) rises rapidly 
from $z=0-1$, such that by $z>0.7$ or so, these systems dominate the 
IR luminosity density. Their fractional contribution remains relatively constant, above 
this point. ULIRGs ($L_{\rm IR}>10^{12}\,L_{\sun}$) 
also rise rapidly in prominence, from negligible 
contributions to the IR luminosity density ($\lesssim1\%$) at 
low redshifts to comparable $\sim20-50\%$ contributions to LIRGs at 
$z\sim2$, and by $z>3$ dominating the total IR emission. 
The contribution from HyLIRGs ($L_{\rm IR}>10^{13}\,L_{\sun}$) 
rises in similar fashion, but is always much less than the contribution from 
ULIRGs -- the highest contributions we find from such systems 
are at $z\gtrsim3$ at $\sim1-5\%$ of the luminosity density. 

These trends simply reflect the predicted (and observed) 
evolution in the IR luminosity function break $\sim L_{\ast}$ (see Figure~\ref{fig:lf.fit.results}) 
-- i.e.\ the fact 
that all active systems become more IR luminous at high redshifts. 
It should be clear both from our discussion in \S~\ref{sec:lfs:transitions}, 
and from direct comparison with Figure~\ref{fig:lumden}, 
that these luminosity classes do {\em not} necessarily represent 
distinct physical object classes. As we have shown, we expect 
most of the LIRGs at $z\sim1$, and ULIRGs at $z>2$, to 
be ``normal'' star forming systems, with their luminosities driven 
by increasing gas content and rapid growth.

\section{Conclusions}
\label{sec:discussion}

We present a simple model for the distribution of 
star formation rates and infrared luminosities owing to ``normal'' star-forming disks, 
merger-induced starbursts, and AGN. 
Comparing this with observations, we find reasonable agreement at $z\sim0-3$. 
At all redshifts, we find that the low-luminosity population 
is dominated by disks, whereas the high-luminosity population 
becomes progressively more dominated by merger-induced 
bursts and then, ultimately, obscured AGN. 

The threshold for this transition is always at high luminosities 
and low space densities. At higher redshifts, gas fractions in all 
systems increase -- hence, specific star formation rates 
even in ``typical'' systems are much higher at high-$z$. As a consequence, 
the luminosity threshold for the disk-merger transition increases with redshift, 
from between LIRGs and ULIRGs in the local Universe, to 
ULIRG luminosities at $z\sim1$, to HyLIRG luminosities at $z\sim2-4$. 
Similarly, the threshold for AGN dominance, at the bright ULIRG range 
at low redshifts, rises to HyLIRG luminosities at $z\sim1-2$ 
and to the very brightest HyLIRGs at $z>2$.

We provide simple fitting functions for each of these quantities, 
and show how they depend on different parameters in the model. 
Most critically, it is the evolution in galaxy gas fractions that drives most of the 
evolution in the IR LFs. Observations have shown that typical gas fractions in 
massive, star-forming galaxies increase rapidly with redshift 
\citep[see e.g.][]{erb:lbg.gasmasses,
bouche:z2.kennicutt,puech:tf.evol,
mannucci:z3.gal.gfs.tf,
forsterschreiber:z2.sf.gal.spectroscopy}. 
This naturally follows from the facts that cooling rates onto galaxies at high 
redshift are much higher than at low redshift, and 
there has simply been less time to process gas into stars 
(higher cosmic densities may also make stellar and AGN feedback 
relatively less efficient). 
A higher gas fraction by a factor of $\sim3$, as implied by observations of 
Milky-Way mass disks at $z\sim2-3$, leads to a factor $\sim5-6$ higher 
star formation rate, according to the Kennicutt-Schmidt relation. At high 
masses/luminosities, i.e.\ where the number density of systems 
is falling exponentially, such a systematic increase in luminosity more 
than offsets the declining space density of massive galaxies with redshift. 

Other parameters have little effect. Changes in galaxy sizes and/or 
structural parameters, while potentially very important for e.g.\ how 
gas fractions are maintained, make little difference to the SFR distributions 
given some gas fraction and stellar mass distributions. 
Allowing for a more steep index in the \citet{kennicutt98} relation 
may help to account for the most 
luminous observed systems such as bright sub-millimeter galaxies 
at the Hyper-LIRG threshold at $z\sim2-4$ \citep{chapman:submm.lfs}. However, 
the number counts of such objects remains quite uncertain, and 
recently it has been suggested that the average counts might be much lower 
with cosmic variance still a concern \citep{austermann:2009.aztec.submm.source.counts}. 
Larger samples and better calibration of bolometric corrections -- in particular, 
real knowledge of the appropriate dust temperatures for conversion to 
total-IR luminosities, which requires sampling both sides of the cold 
dust peak \citep[see e.g.][]{younger:mm.obs.z2.ulirgs} -- 
will be needed for better understanding of extreme systems. 

Our simple model succeeds reasonably well at explaining the observed 
global SFR density. 
At all redshifts, normal systems dominate the global SFR density. 
Obscured AGN contribute little, $\lesssim5\%$, to the total IR luminosity density. 
Substantial bias to IR-based SFR density estimates from obscured AGN 
would require an undiscovered population of heavily, isotropically 
obscured sources with luminosity densities $\sim5$ times what is 
currently suggested (which would be in conflict with relic BH mass densities). 

The contribution of merger-induced bursts is similarly small, 
$\sim5-10\%$ at most redshifts.
This owes both to the physics discussed above -- disks also rapidly 
increase their SFR density with redshift owing to higher gas 
fractions -- and also to the fact 
that increasing disk gas fractions arbitrarily will not 
continue to increase the merger-induced burst contribution 
arbitrarily. Rather, as discussed in detail in \citet{hopkins:disk.survival,hopkins:disk.survival.cosmo}, 
at high gas fractions angular momentum loss in mergers becomes less efficient -- 
thus for an otherwise identical merger with a much larger gas fraction, the 
fraction of gas funneled into the nuclear starburst, relative to the total available, 
will be less (and the fraction that remains in an extended disk distribution to continue 
``normal'' mode star formation will be larger), even if the absolute mass in the burst 
is larger. Similarly, at a given mass, the distribution of SFRs at all redshifts is dominated 
by normal-mode star formation -- merger-induced bursts are important only in 
the high-SFR tail ($\sim2-3\,\sigma$) of the distribution. 
These trends explain a number of recent observations 
that similarly indicate a small effect of merger enhancements 
to star formation rates, and that show that most star formation by number and 
luminosity density appears to follow a simple trend or ``main sequence'' 
as a function of galaxy stellar mass and redshift 
\citep{blain:ir.lf.synthesis.model,
noeske:2007.sfh.part1,
noeske:sfh,papovich:ssfr,
bell:morphology.vs.sfr,
jogee:merger.density.08,
robaina:2009.sf.fraction.from.mergers}.

Support for these fractions also comes from completely 
independent sources. Recently, a number of high-resolution studies of 
spheroid formation via galaxy mergers have shown that 
properties such as the surface brightness profiles, 
sizes, concentrations, kinematics, and isophotal shapes of 
spheroids are very sensitive to the mass fractions formed 
in such bursts, which produce dense, disky, nuclear mass concentrations, 
versus the mass in a more extended envelope formed via the violent 
relaxation of the pre-burst stellar disks \citep[see e.g.][]{cox:kinematics,
naab:gas,robertson:fp,burkert:anisotropy,
jesseit:merger.rem.spin.vs.gas,hopkins:cusps.mergers,
hopkins:cusps.fp,hopkins:cusps.evol}. 
In particular, typical $\sim L_{\ast}$ 
early-type galaxies, which dominate the spheroid stellar 
mass density, have properties that are reproduced accurately by 
simulations if and only if this burst fraction is $\sim10\%$ 
\citep[for details, see][]{hopkins:cusps.ell,hopkins:cores}. Independent analysis of their 
stellar population properties leads to similar conclusions 
\citep{mcdermid:sauron.profiles,sanchezblazquez:ssp.gradients,
reda:ssp.gradients,foster:metallicity.gradients.prep}. 

There is an important technical distinction between the total SFR density in 
``ongoing'' or recent mergers and that actually {\em induced} by the merger (we present 
predictions for both). 
The former includes systems in their ``normal'' star-forming mode, observable for $\sim$Gyr 
as perturbed or in pairs; the latter reflects specifically the $\sim10^{8}\,$yr event where 
gravitational torques drive a nuclear starburst. 
Under some circumstances, 
especially in very gas-rich mergers, the sum of this ``normal mode'' 
star formation over a long $\sim$Gyr duration yields 
significantly more total stellar mass formed than in the burst itself. 
The ``ongoing'' merger SFR density must, of course, rise with the observed 
merger fraction, reaching $\sim20\%$ at $z=2$ and as high as $\sim20-50\%$ 
at $z>4$; however we stress that these high fractions 
reflect predominantly the ``normal'' modes of star formation 
simply present in systems that may be on their way to merging. 

Finally, 
we caution that the above comparisons are approximate, and intended as a broad 
means of comparing the primary drivers of star formation and their contributions 
relative to observed IR luminosity functions and SFR distributions. 
We have ignored a number of 
potentially important effects: for example, obscuration is a strong function of time 
in a merger, and may affect various luminosities and morphological stages 
differently. Moreover, our simple linear addition of the star formation contribution 
of mergers to the IR LF and the AGN contribution is only technically correct 
if one or the other dominates the IR luminosity at a given time in the merger; however, 
there are clearly times during the final merger stages when the contributions 
are comparable. Resolving these issues requires detailed, time-dependent 
radiative transfer solutions through high-resolution simulations that properly 
sample the merger and quiescent galaxy parameter space at each redshift, 
and is outside the scope of this work \citep[although an important subject for future, 
more detailed study; see, e.g.][]{li:radiative.transfer,narayanan:smg.modeling,
narayanan:molecular.gas.in.smgs,younger:warm.ulirg.evol}.
It would be a mistake, therefore, to read too much into 
e.g.\ the detailed predictions for sub-millimeter galaxies or other extreme 
populations in Figure~\ref{fig:ir.lfs} 
that may have complex dust geometries and/or a non-trivial 
mix of contributions from all of ``normal'' and ``burst'' mode star formation as 
well as AGN. However, most of our 
predicted qualitative trends, including the evolution of the luminosity density 
(and approximate relative contribution of mergers) and the shift in where 
quiescent or merger-driven populations dominate the bright IR LF, should 
be robust.

\acknowledgments 
We thank Eliot Quataert, Lin Yan, 
Dave Sanders, Nick Scoville, 
and Kevin Bundy for helpful 
discussions throughout the development of this manuscript. 
Support for PFH was provided by the Miller Institute for Basic Research 
in Science, University of California Berkeley.
JDY acknowledges support from NASA through Hubble Fellowship grant
\#HF-51266.01 awarded by the Space Telescope Science Institute,which
is operated by the Association of Universities for Research in
Astronomy, Inc., for NASA, under contract NAS 5-26555.
Many of the computations in this paper were run on the Odyssey cluster
supported by the FAS Research Computing Group at Harvard University.
\\

\bibliography{/Users/phopkins/Documents/lars_galaxies/papers/ms}

\clearpage

\begin{footnotesize}
\begin{landscape}
\ctable[
  caption={{\normalsize Fits to Model IR LF Predictions}\label{tbl:fits}},center
  ]{lcccccccccccc}{
\tnote[a]{Refers to sub-sample of objects for which the 
fit pertains.}
\tnote[b]{Parameters of best fit to the redshift-dependent 
form of the IR LF (Equations~\ref{eqn:double.pwr.law}-\ref{eqn:modschechter}). 
$L_{0}$ is the break luminosity $L_{\ast}$ at $z=0$, in $\log{(L_{\rm 0}/L_{\sun})}$.}
\tnote[c-d]{Redshift dependence of the break luminosity $L_{\ast}$, per 
Equation~\ref{eqn:param.z.evol} 
($\log{\{L_{\ast}/L_{\sun}\}} = L_{0} + L^{\prime}\,\xi + L^{\prime\prime}\,\xi^{2}$, 
where $\xi\equiv \log{(1+z)}$.)}
\tnote[e]{Log LF normalization $\phi_{\ast}$ at $z=0$, in 
${\rm Mpc^{-3}\,\log^{-1}{L_{\rm IR}}}$.}
\tnote[f-g]{Dependence of normalization $\phi_{\ast}$ on redshift 
($\log{\{\phi_{\ast}/{\rm Mpc^{-3}\,\log^{-1}{L_{\rm IR}}}\}} = 
\phi_{0} + \phi^{\prime}\,\xi + \phi^{\prime\prime}\,\xi^{2}$).}
\tnote[h]{Faint-end IR LF slope $\alpha$ at $z=0$.}
\tnote[i-j]{Dependence of faint-end slope $\alpha$ on redshift
($\alpha = \alpha_{0} + \alpha^{\prime}\,\xi + \alpha^{\prime\prime}\,\xi^{2}$).}
\tnote[k]{Bright-end IR LF slope $\beta$ at $z=0$.}
\tnote[l-m]{Dependence of bright-end slope $\beta$ on redshift 
($\beta = \beta_{0} + \beta^{\prime}\,\xi + \beta^{\prime\prime}\,\xi^{2}$).\\ }
\tnote[ ]{Observed galaxy mass functions extend to $z\approx4$, the 
range used for these fits. Extrapolations beyond this redshift should be 
considered with caution. \\
If this is done, however, a minimum should 
be imposed on the bright-end slope at $\beta\gtrsim2$. }
}{
\hline\hline
\multicolumn{1}{c}{Object Class\tmark[\ a]} &
\multicolumn{1}{c}{$L_{0}(\pm \Delta L_{0})$\tmark[\ b]} &
\multicolumn{1}{c}{$L^{\prime}$\tmark[\ c]} & 
\multicolumn{1}{c}{$L^{\prime\prime}$\tmark[\ d]} & 
\multicolumn{1}{c}{$\phi_{0}$\tmark[\ e]} & 
\multicolumn{1}{c}{$\phi^{\prime}$\tmark[\ f]} & 
\multicolumn{1}{c}{$\phi^{\prime\prime}$\tmark[\ g]} & 
\multicolumn{1}{c}{$\alpha_{0}$\tmark[\ h]} & 
\multicolumn{1}{c}{$\alpha^{\prime}$\tmark[\ i]} & 
\multicolumn{1}{c}{$\alpha^{\prime\prime}$\tmark[\ j]} & 
\multicolumn{1}{c}{$\beta_{0}$\tmark[\ k]} & 
\multicolumn{1}{c}{$\beta^{\prime}$\tmark[\ l]}  & 
\multicolumn{1}{c}{$\beta^{\prime\prime}$\tmark[\ m]}  \\
\hline
\multicolumn{13}{c}{Double Power-Law Fit (Full) - Standard Model}\\
\hline
{\rm Normal/Star-Forming} 
& 11.37(0.14) & 2.17(0.82) & -1.15(1.13) 
& -3.97(0.50) & 5.27(2.81) & -6.11(3.53)
& 1.57(0.45) & -3.40(2.46) & 2.09(3.06) 
& 6.60(1.33) & 7.46(8.12) & -17.57(10.81)\\
{\rm Merger/Burst} 
& 12.16(0.26) & 1.93(1.53) & -2.41(2.09)
& -6.35(0.52) & 5.15(3.06) & -4.77(4.02)
& 1.13(0.21) & -0.86(1.30) & 0.30(1.76)
& 3.23(0.59) & 3.94(3.55) & -7.42(4.69)\\
{\rm Obscured AGN} 
& 12.15(0.29) & 2.07(1.63) & -3.19(2.10)
& -6.62(0.47) & 5.69(2.56) & -5.39(3.15) 
& 0.73(0.23) & -0.64(1.37) & -0.83(1.85) 
& 2.55(0.50) & 3.90(2.91) & -7.41(3.74)\\
\hline
\multicolumn{13}{c}{Double Power-Law Fit (First-Order in $z$) - Standard Model}\\
\hline
{\rm Normal/Star-Forming} & 11.31(0.12) & 1.99(0.29) & 0 & -3.30(0.41) & 0.16(0.88) & 0 & 1.39(0.35) & -1.68(0.72) & 0 & 6.59(1.10) & 0.95(3.44) & 0\\
{\rm Merger/Burst} & 12.08(0.30) & 1.23(0.69) & 0 & -5.67(0.59) & 0.72(1.33) & 0 & 1.07(0.25) & -0.53(0.55) & 0 & 3.18(0.60) & 0.79(1.71) & 0\\
{\rm Obscured AGN} & 11.95(0.34) & 1.37(0.75) & 0 & -5.80(0.53) & 0.76(1.11) & 0 & 0.69(0.28) & -0.83(0.60) & 0 & 2.40(0.45) & 1.06(1.32) & 0\\
\hline
\multicolumn{13}{c}{Modified Schechter Function Fit - Standard Model}\\
\hline
{\rm Normal/Star-Forming} & 10.46(0.58) & 2.35(0.97) & 0 & -1.64(0.70) & -1.28(1.14) & 0 & 0.46(0.74) & -0.99(1.24) & 0 & 0.73(0.22) & 0.25(0.40) & 0 \\
{\rm Merger/Burst} & 10.23(1.83) & 1.05(0.64) & 0 & -3.41(1.10) & 0.21(0.57) & 0 & 0.28(0.72) & -0.26(0.76)  & 0 & 0.36(0.15) & 0.0(0.0) & 0 \\
{\rm Obscured AGN} & 10.10(0.66) & 0.90(0.60) & 0 & -4.25(0.40) & -0.36(0.81) & 0 & 0.0(0.0) & -0.60(0.61) & 0 & 0.33(0.06) & 0.0(0.0) & 0 \\
\hline\hline
\\
\hline
\multicolumn{13}{c}{Double Power-Law Fit (Full) - Steeper Kennicutt-Schmidt Index}\\
\hline
{\rm Normal/Star-Forming} 
& 11.28(0.14) & 2.43(0.84) & -1.12(1.14) 
& -4.09(0.53) & 5.67(2.91) & -6.35(3.58)
& 1.69(0.48) & -3.93(2.55) & 2.48(3.09) 
& 6.57(1.36) & 7.39(8.12) & -17.75(10.66)\\
{\rm Merger/Burst} 
& 12.39(0.25) & 2.06(1.49) & -2.59(2.07)
& -6.61(0.45) & 5.00(2.65) & -4.84(3.52)
& 0.96(0.17) & -0.85(1.05) & 0.48(1.43)
& 2.93(0.50) & 3.71(3.08) & -6.88(4.09)\\
{\rm Obscured AGN} 
& 12.15(0.29) & 2.07(1.63) & -3.19(2.10)
& -6.62(0.47) & 5.69(2.56) & -5.39(3.15) 
& 0.73(0.23) & -0.64(1.37) & -0.83(1.85) 
& 2.55(0.50) & 3.90(2.91) & -7.41(3.74)\\
\hline
\multicolumn{13}{c}{Modified Schechter Function Fit - Steeper Kennicutt-Schmidt Index}\\
\hline
{\rm Normal/Star-Forming} 
& 10.34(0.56) & 2.79(0.91) & 0 
& -1.62(0.69) & -1.34(1.11) & 0 
& 0.51(0.71) & -1.07(1.16) & 0 
& 0.72(0.21) & 0.27(0.38) & 0 \\
{\rm Merger/Burst} 
& 10.62(1.53) & 1.05(0.59) & 0 
& -3.83(0.95) & 0.30(0.51) & 0 
& 0.27(0.53) & -0.21(0.59)  & 0 
& 0.36(0.13) & 0.0(0.0) & 0 \\
{\rm Obscured AGN} 
& 10.10(0.66) & 0.90(0.60) & 0 
& -4.25(0.40) & -0.36(0.81) & 0 
& 0.0(0.0) & -0.60(0.61) & 0 
& 0.33(0.06) & 0.0(0.0) & 0 \\
\hline\hline
\\
\hline
\multicolumn{13}{c}{Double Power-Law Fit (Full) - No Gas Fraction Evolution}\\
\hline
{\rm Normal/Star-Forming} 
& 11.38(0.15) & 1.06(0.96) & -1.50(1.41) 
& -4.00(0.57) & 4.36(3.45) & -5.71(4.65)
& 1.57(0.53) & -2.12(3.36) & 1.05(4.67) 
& 6.70(1.52) & 7.34(9.66) & -16.84(13.18)\\
{\rm Merger/Burst} 
& 12.20(0.26) & 1.35(1.54) & -2.77(2.12)
& -6.40(0.53) & 4.51(3.18) & -3.53(4.26)
& 1.12(0.22) & -0.21(1.42) & -0.58(2.05)
& 3.33(0.64) & 4.08(3.85) & -7.61(5.03)\\
{\rm Obscured AGN} 
& 12.15(0.33) & 1.91(1.92) & -2.88(2.62)
& -6.60(0.51) & 5.45(2.99) & -5.29(3.90) 
& 0.75(0.24) & -0.83(1.48) &  0.33(2.04) 
& 2.54(0.53) & 3.65(3.20) & -6.66(4.23)\\
\hline
\multicolumn{13}{c}{Modified Schechter Function Fit - No Gas Fraction Evolution}\\
\hline
{\rm Normal/Star-Forming} 
& 10.37(1.08) & 0.93(1.74) & 0 
& -1.52(0.94) & -1.55(1.47) & 0 
& 0.27(1.48) & -1.01(2.53) & 0 
& 0.69(0.34) & 0.18(0.61) & 0 \\
{\rm Merger/Burst} 
& 10.09(2.41) & 0.18(0.74) & 0 
& -3.41(1.12) & 0.58(0.61) & 0 
& 0.18(0.92) & -0.15(0.94)  & 0 
& 0.35(0.19) & 0.0(0.0) & 0 \\
{\rm Obscured AGN} 
& 10.07(0.71) & 0.73(0.68) & 0 
& -4.31(0.41) & 0.38(0.61) & 0 
& 0.0(0.0) & -0.28(0.65) & 0 
& 0.32(0.06) & 0.0(0.0) & 0 \\
\hline\hline\\
}
\end{landscape}
\end{footnotesize}

\clearpage

\begin{appendix}

\section{Predicted Luminosity Functions as a Function of Wavelength}
\label{sec:appendix:multiwavelength}

\begin{figure*}
    \centering
    \plotside{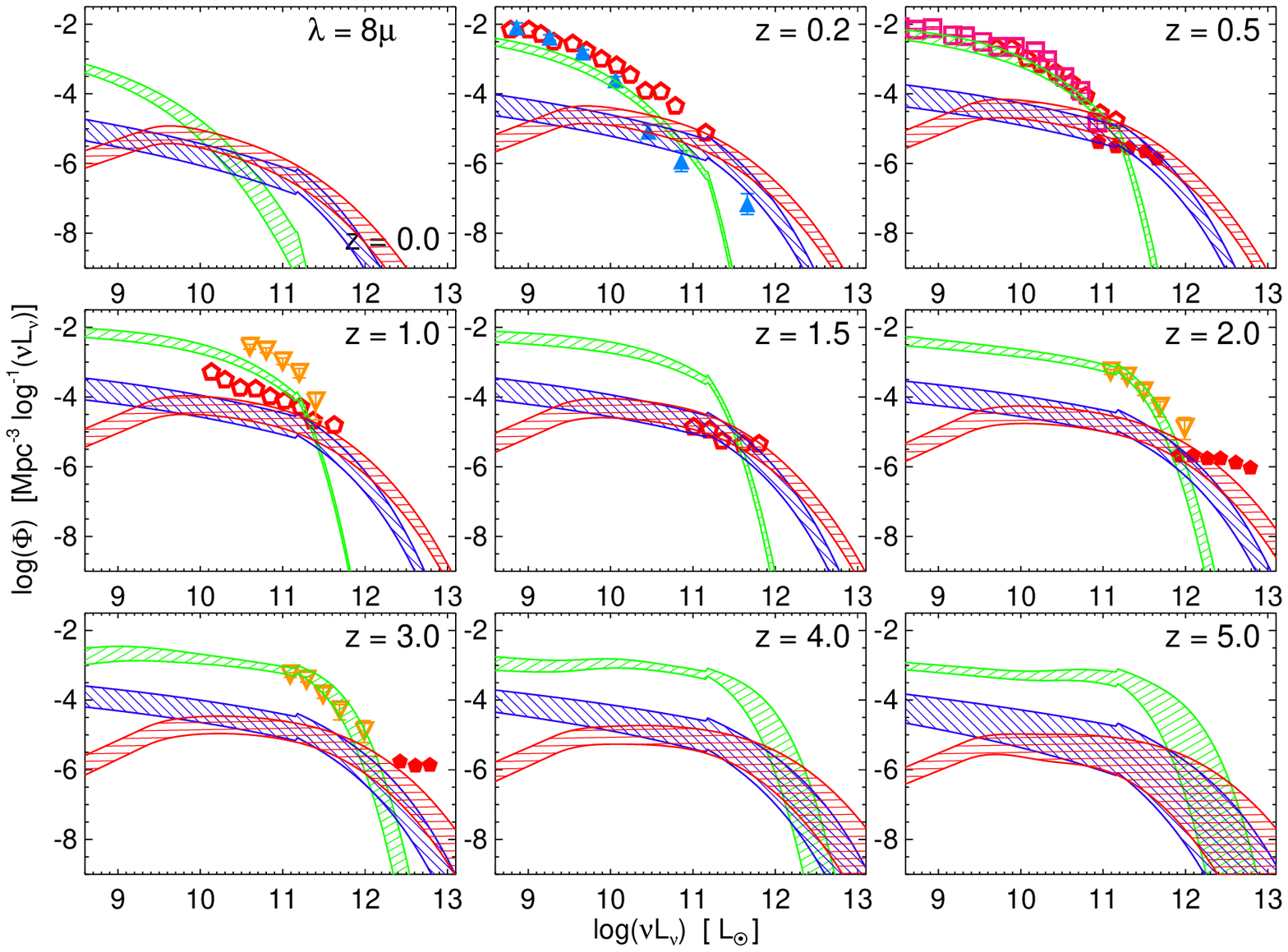}
    \caption{As Figure~\ref{fig:ir.lfs}, but for the luminosity function in a specific 
    rest-frame band (here, $\lambda=8\,\mu$) as a function of redshift. 
    We stress that we do {\em not} model the SEDs {\em a priori}, but simply 
    adopt a specific set of empirical templates -- as such, the information 
    in this plot is identical to that in Figure~\ref{fig:ir.lfs}. We compare to the 
    same observations as Figure~\ref{fig:ir.lfs}, for the observations at or 
    near this rest-frame wavelength. Solid red pentagons show the estimates 
    from \citet{babbedge:swire.lfs} specifically for the contribution of AGN 
    at this wavelength. 
    \label{fig:ir.lfs.8m}}
\end{figure*}

\begin{figure*}
    \centering
    \plotside{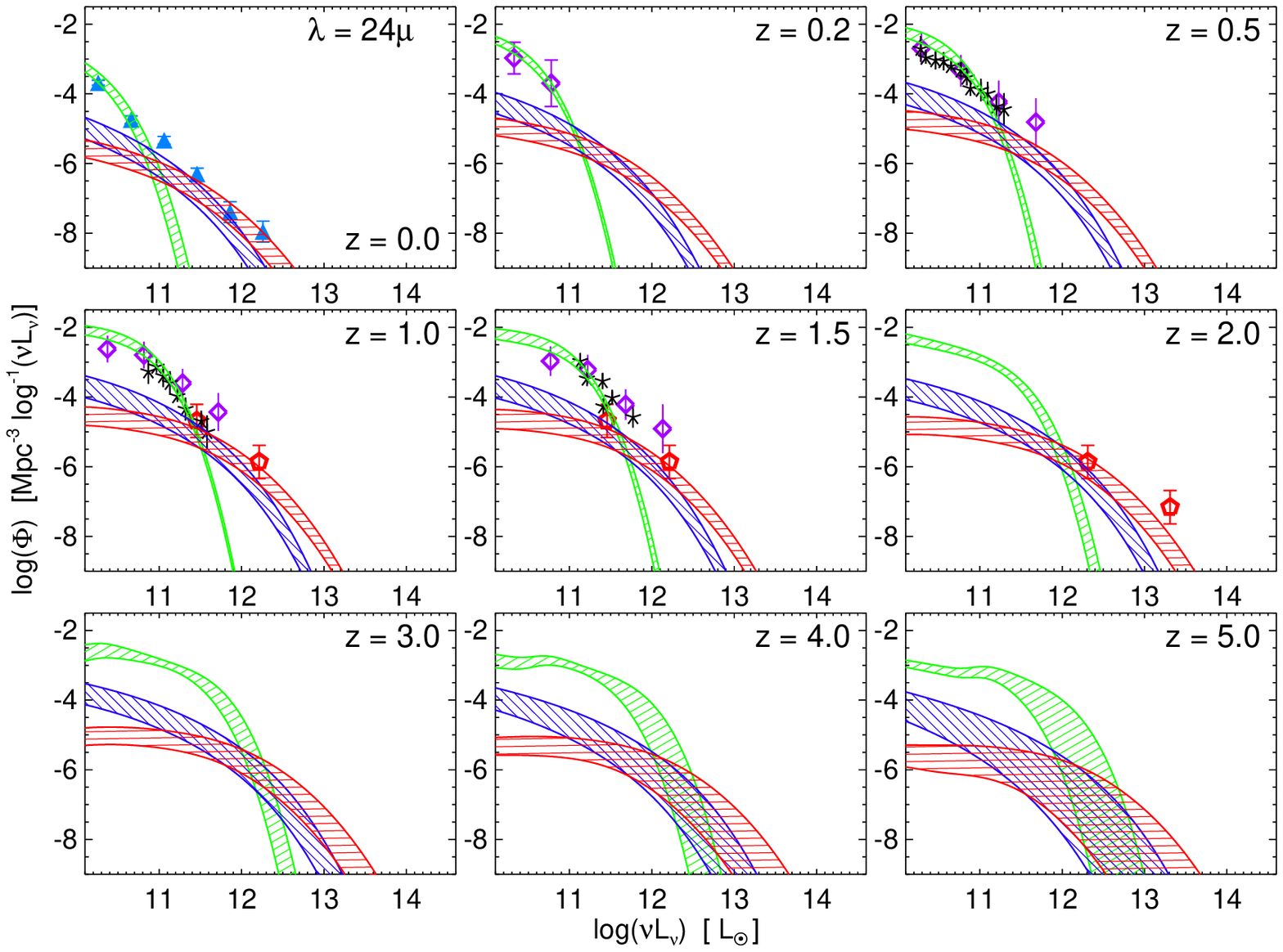}
    \caption{As Figure~\ref{fig:ir.lfs.8m}, but at rest-frame $24\,\mu$. 
    We compare observations spanning rest-frame $15-35\,\mu$ 
    (corrected with the same standard bolometric corrections to $24\,\mu$). 
    \label{fig:ir.lfs.24m}}
\end{figure*}

\begin{figure*}
    \centering
    \plotside{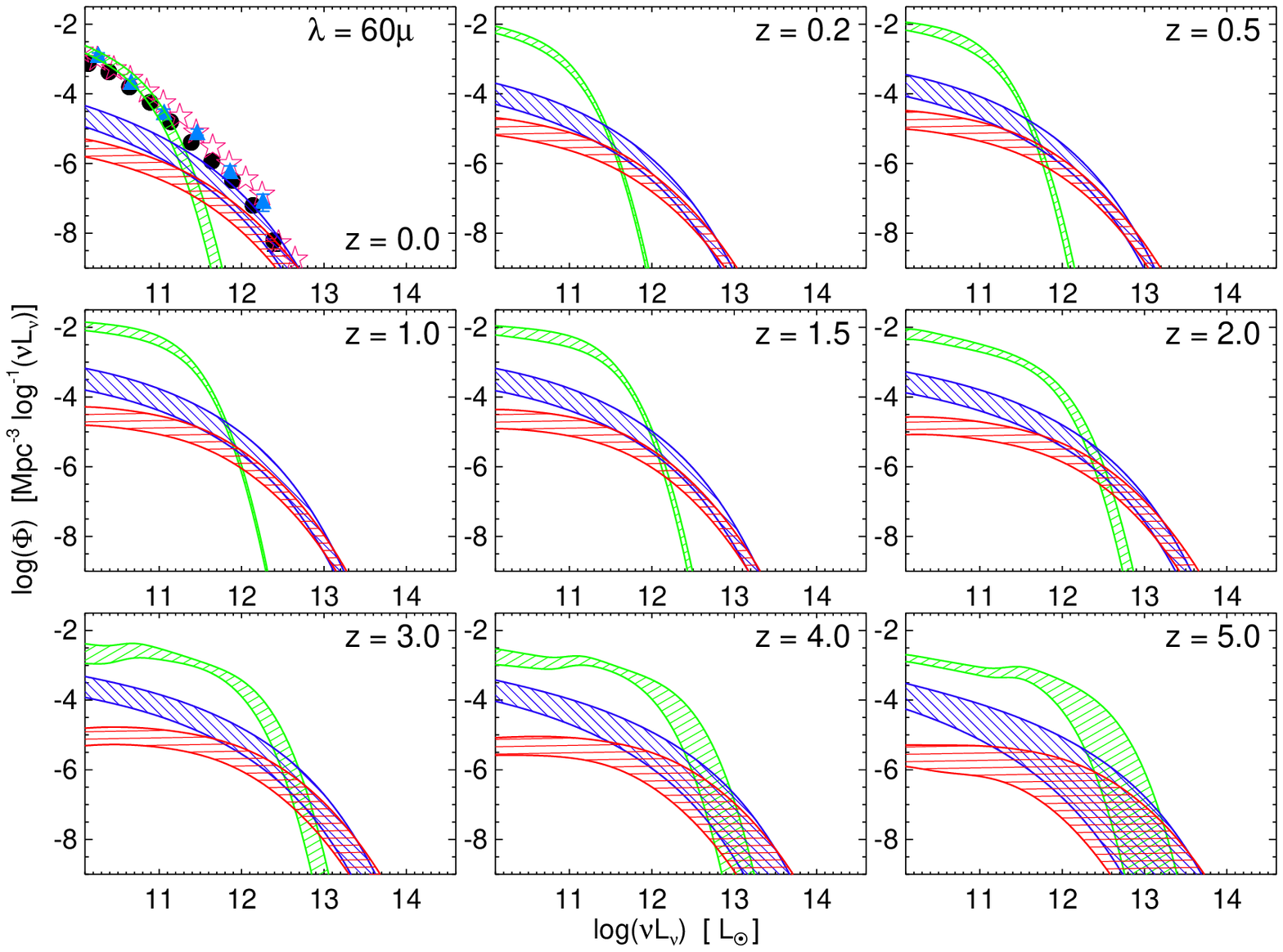}
    \caption{As Figure~\ref{fig:ir.lfs.8m}, but at rest-frame $60\,\mu$. 
    We compare observations spanning rest-frame $60-70\,\mu$ 
    (corrected with the same standard bolometric corrections to $60\,\mu$). 
    \label{fig:ir.lfs.60m}}
\end{figure*}

\begin{figure*}
    \centering
    \plotside{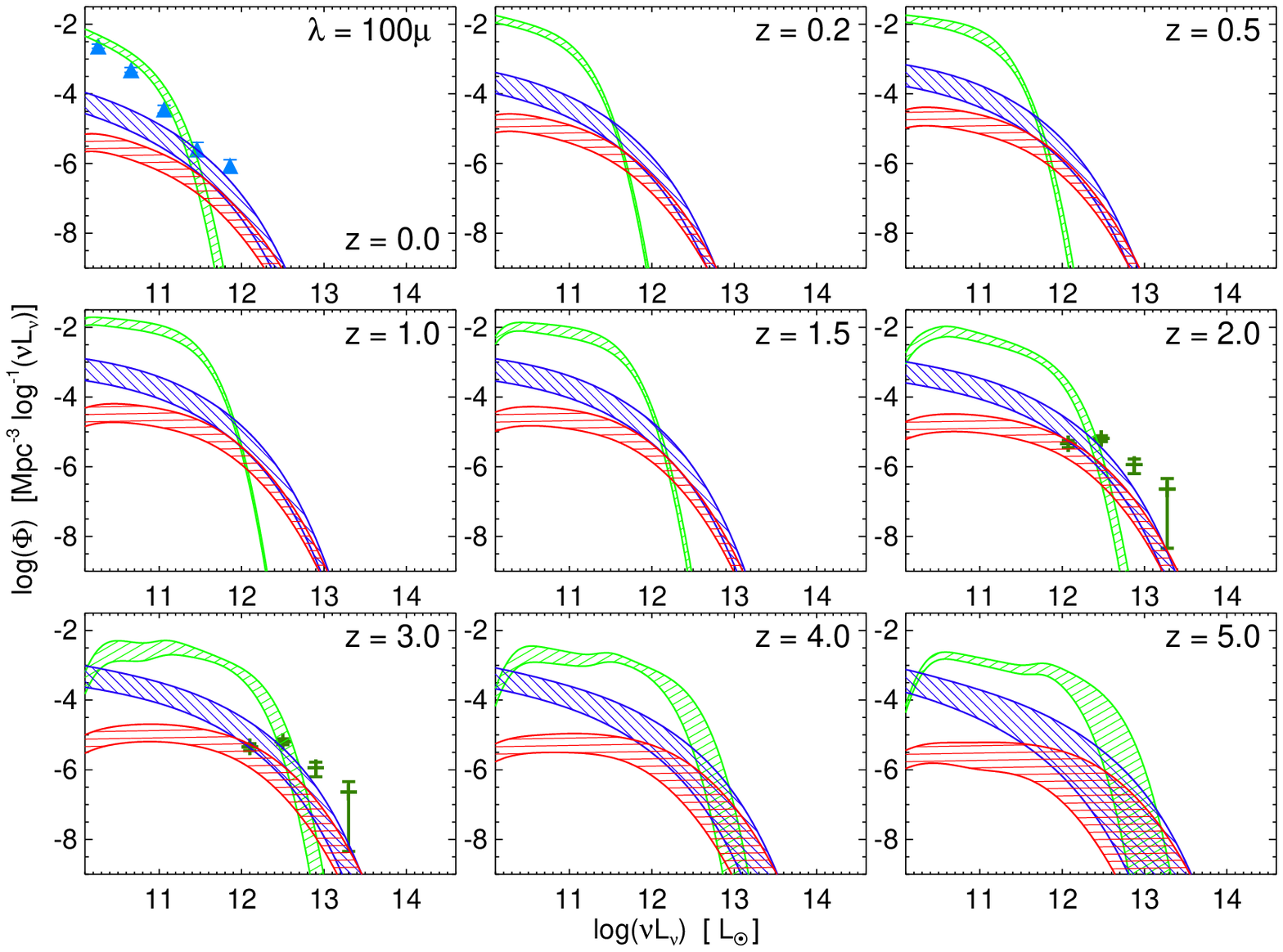}
    \caption{As Figure~\ref{fig:ir.lfs.8m}, but at rest-frame $100\,\mu$. 
    We compare observations spanning rest-frame $80-120\,\mu$ 
    (corrected with the same standard bolometric corrections to $100\,\mu$). 
    \label{fig:ir.lfs.100m}}
\end{figure*}

\begin{figure*}
    \centering
    \plotside{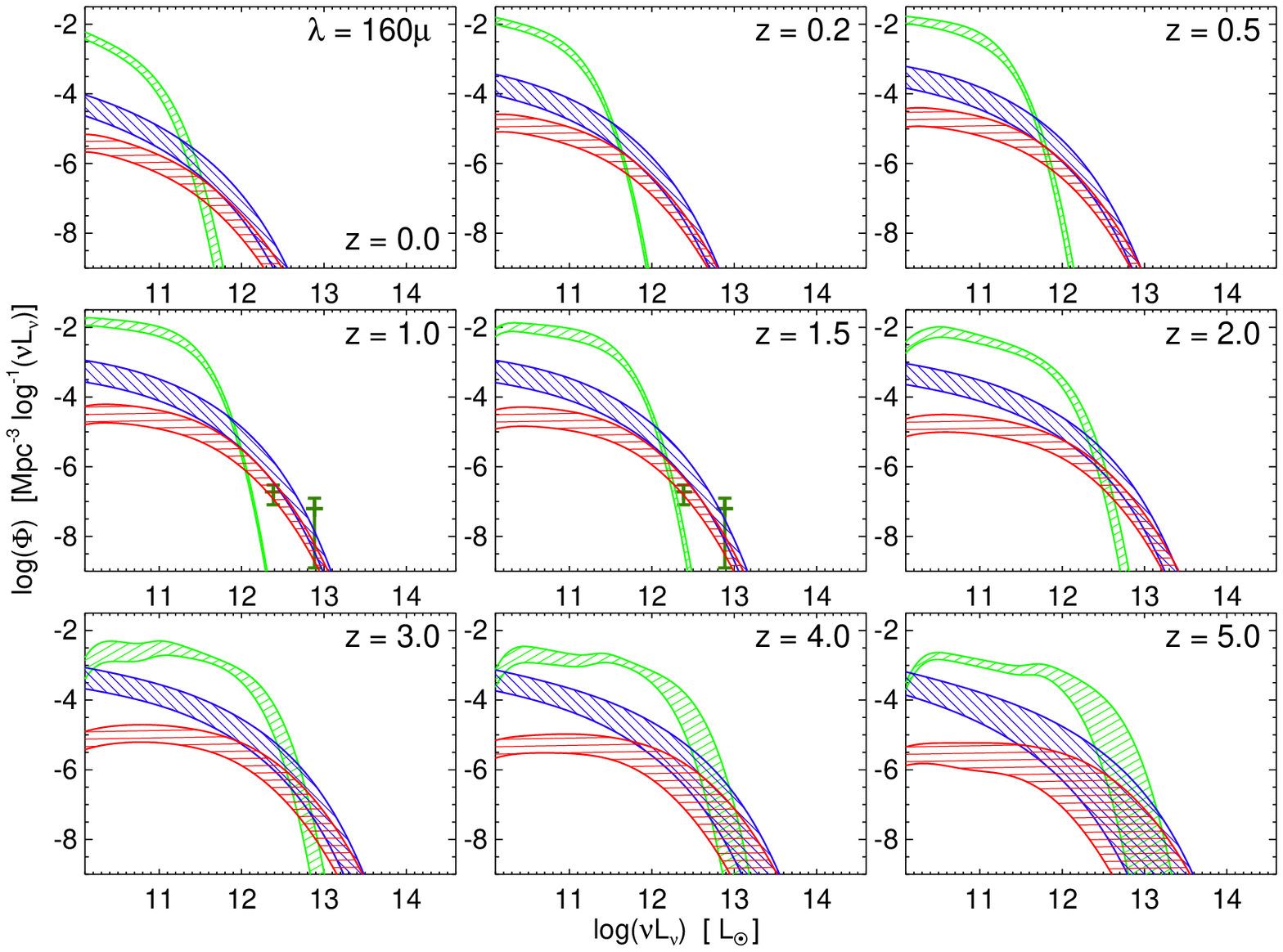}
    \caption{As Figure~\ref{fig:ir.lfs.8m}, but at rest-frame $160\,\mu$. 
    We compare observations spanning rest-frame $120-350\,\mu$ 
    (corrected with the same standard bolometric corrections to $160\,\mu$). 
    \label{fig:ir.lfs.160m}}
\end{figure*}

Figure~\ref{fig:ir.lfs} presents the predicted total IR luminosity functions from 
the models discussed here. To facilitate comparison with observations (and broaden 
the range of observations to which we can compare), we here present 
corresponding predictions in a number of different rest-frame wavelengths. 
Figures~\ref{fig:ir.lfs.8m}, \ref{fig:ir.lfs.24m}, 
\ref{fig:ir.lfs.60m}, \ref{fig:ir.lfs.100m}, 
\&\ \ref{fig:ir.lfs.160m} present the predicted LFs at rest-frame wavelengths of 
$8\,\mu$, $24\,\mu$, $60\,\mu$, $100\,\mu$, and $160\,\mu$, respectively. 
For each, we compare to the available observations at or near that wavelength. 

For the star-forming systems (normal galaxies and mergers), we simply convert our 
predicted SFR and corresponding total IR luminosity to an observed luminosity 
in the given band, given the SED templates (themselves a function of 
bolometric luminosity) discussed in \S~\ref{sec:lfs:pred}, 
namely those from \citet{valiante:ir.lfs.and.numbercounts}, using the model 
SEDs in \citet{dalehelou:ir.sed.templates}. As discussed in \S~\ref{sec:lfs}, 
varying the exact scaling of these corrections within observational uncertainties 
is comparable to the uncertainty from adopting different mass function estimators. 
For the AGN, we adopt the template SEDs for obscured and unobscured systems 
from \citet{hopkins:bol.qlf}, but adopting alternative different template obscured 
or unobscured AGN spectra \citep[e.g.\ those in][]{elvis:atlas,
zakamska:multiwavelength.type.2.quasars,polletta:obscured.qsos,
richards:seds} makes little difference. 

We stress that these predictions should be regarded with considerable caution. 
The models here (or, for that matter, in any fully cosmological model for 
disk/merger/AGN systems as a function of redshift) do {\em not} predict 
full SEDs. Rather, the robust quantity is some more physical number such as the 
total star formation rate (or AGN luminosity and obscured fraction). This 
allows robust estimatiions of total IR luminosity, but we are now using 
a specific, simple empirical conversion between bolometric luminosity 
and luminosity at a given wavelength. If structural properties of galaxies, spatial distributions of 
gas and star formation, dust properties (gas-to-dust ratios, dust spatial distributions and 
clumpiness), clumping factors, and AGN contributions evolve, then these 
conversions will be problematic and may introduce systematic errors. 
In fact, it is very likely that these parameters that govern the SED do, in fact, evolve with redshift, or 
are different in merging and non-merging systems (given the different 
spatial distributions of gas and dust), and/or are a function of the relative AGN/star formation 
balance in the galaxy. 

More detailed modeling, including full, self-consistent radiative transfer treatment of 
high-resolution hydrodynamic simulations of mergers and normal galaxies with 
all the effects above included, will be necessary to predict e.g.\ the distribution of dust 
temperatures and other quantities critical, especially, for comparison with the 
number counts at long wavelengths (e.g.\ sub-millimeter galaxies). 
These models will be presented in future work (in preparation); therefore we 
do not construct such comparisons (or attempt to compile predicted 
number counts) here. 

Nevertheless, the comparisons in Figures~\ref{fig:ir.lfs.8m}-\ref{fig:ir.lfs.160m} are informative, 
and useful for future comparisons with observations, provided appropriate caution is used. 
In particular, this allows us to see how the relative contributions of AGN and star formation 
vary as a function of IR wavelength (as the dust temperatures and SEDs are not the same). 
At shorter wavelengths, e.g.\ $8\,\mu$, AGN play a role at even moderate luminosities. 
We can compare to some observational studies, for example that in \citet{babbedge:swire.lfs}, 
that explicitly separate the AGN and star-forming populations, and find good agreement. 
On the other hand, at the longest wavelengths, the warmer dust temperatures typical in 
AGN lead to their being relatively less important.

\clearpage

\section{Consequences of Model Assumptions}
\label{sec:appendix:assumptions}

\begin{figure}
    \centering
    \plotone{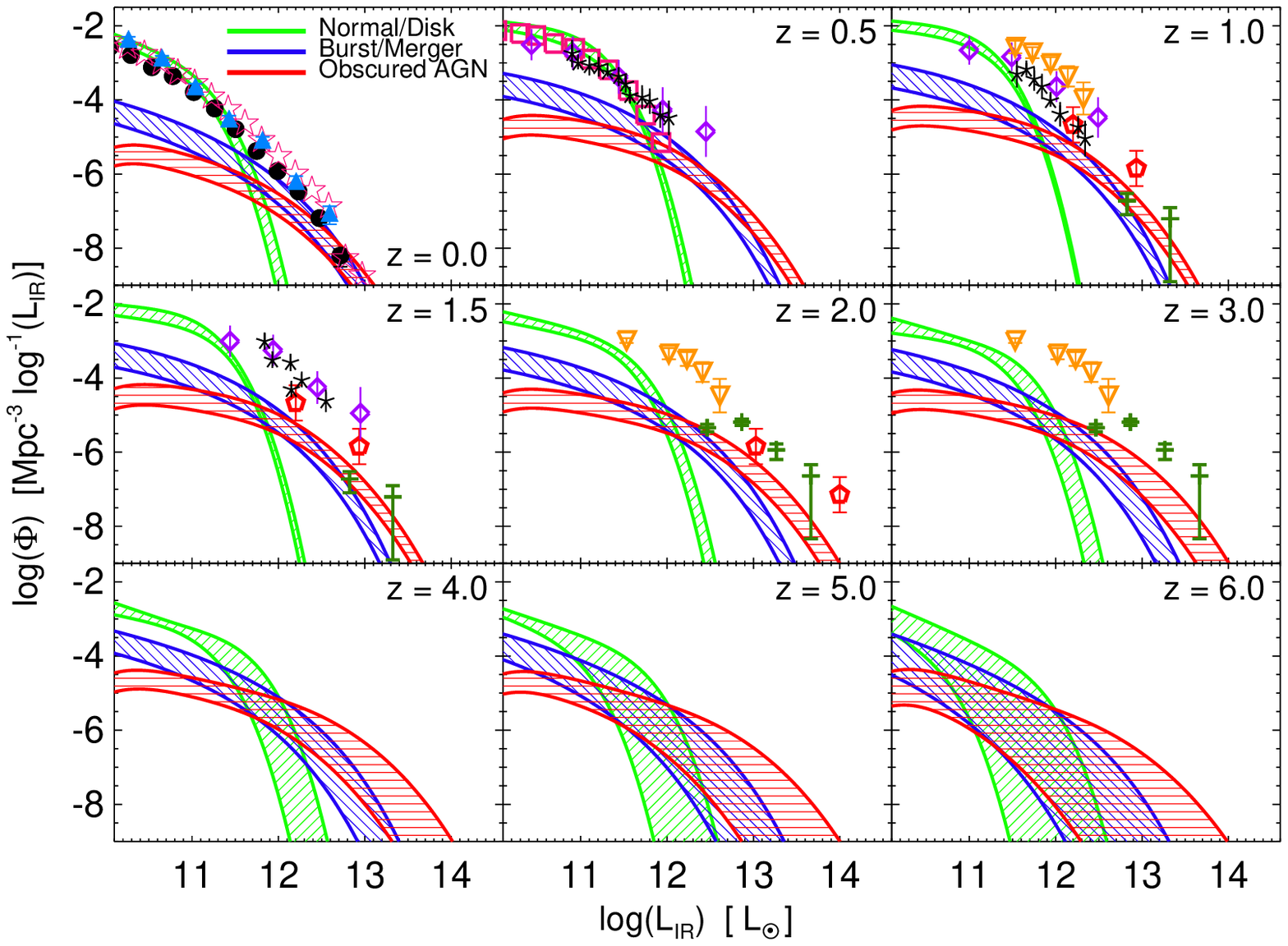}
    \caption{As Figure~\ref{fig:ir.lfs}, but neglecting the 
    increase in galaxy gas fractions with redshift. The SFRs of even normal, 
    low-mass undisturbed disks are significantly under-predicted. 
    \label{fig:ir.lfs.nogasevol}}
\end{figure}
\begin{figure}
    \centering
    \plotone{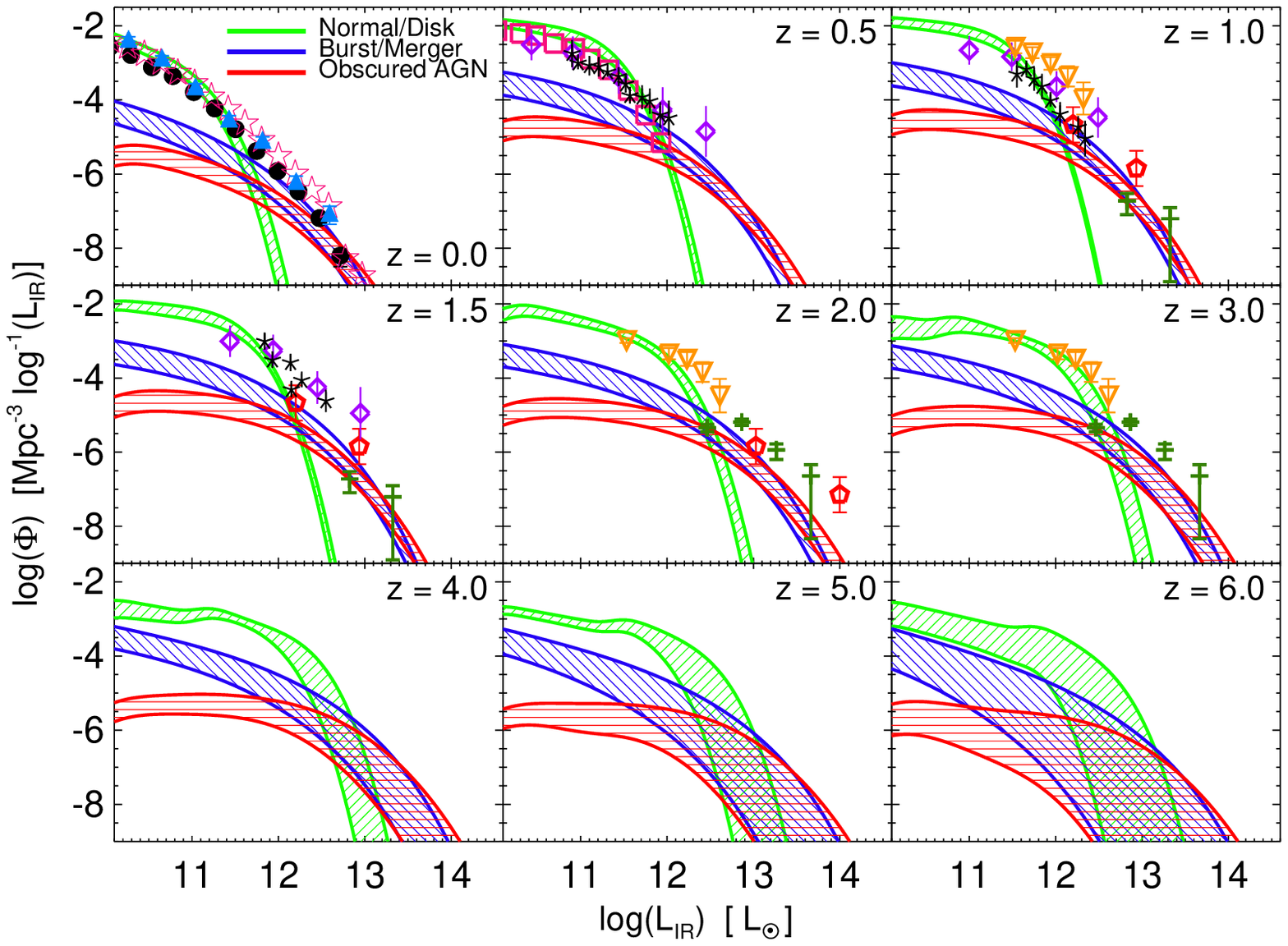}
    \caption{As Figure~\ref{fig:ir.lfs}, but neglecting the evolution in disk 
    sizes with redshift. Because the observed size evolution (of star-forming galaxies) is relative weak 
    ($R_{e}(M_{\ast})\propto(1+z)^{-(0-0.6)}$) and the size evolution (at otherwise fixed properties) 
    only enters into the SFR at sub-linear order, the difference is relatively small (factor $\sim2$). 
    \label{fig:ir.lfs.nosizeevol}}
\end{figure}
\begin{figure}
    \centering
    \plotone{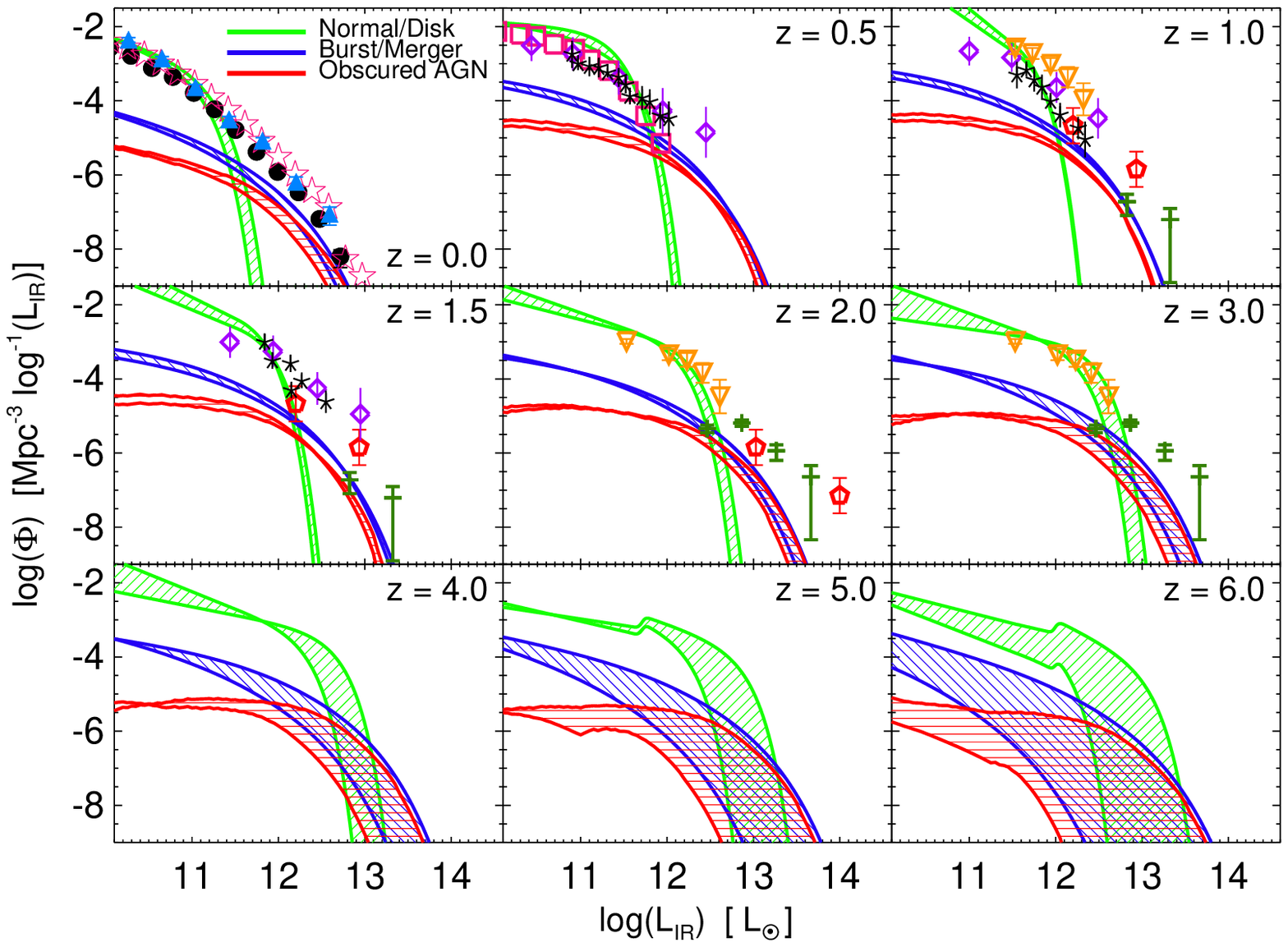}
    \caption{As Figure~\ref{fig:ir.lfs}, but not allowing for any scatter in SFR at fixed 
    galaxy mass, in mergers or disks (and no scatter in obscured fractions/bolometric corrections 
    in AGN). The high-$L$ tail is significantly suppressed. Note that the ``kinks'' at high 
    redshift are artifacts of the analytic fitting functions used. 
    \label{fig:ir.lfs.noscatter}}
\end{figure}
\begin{figure}
    \centering
    \plotone{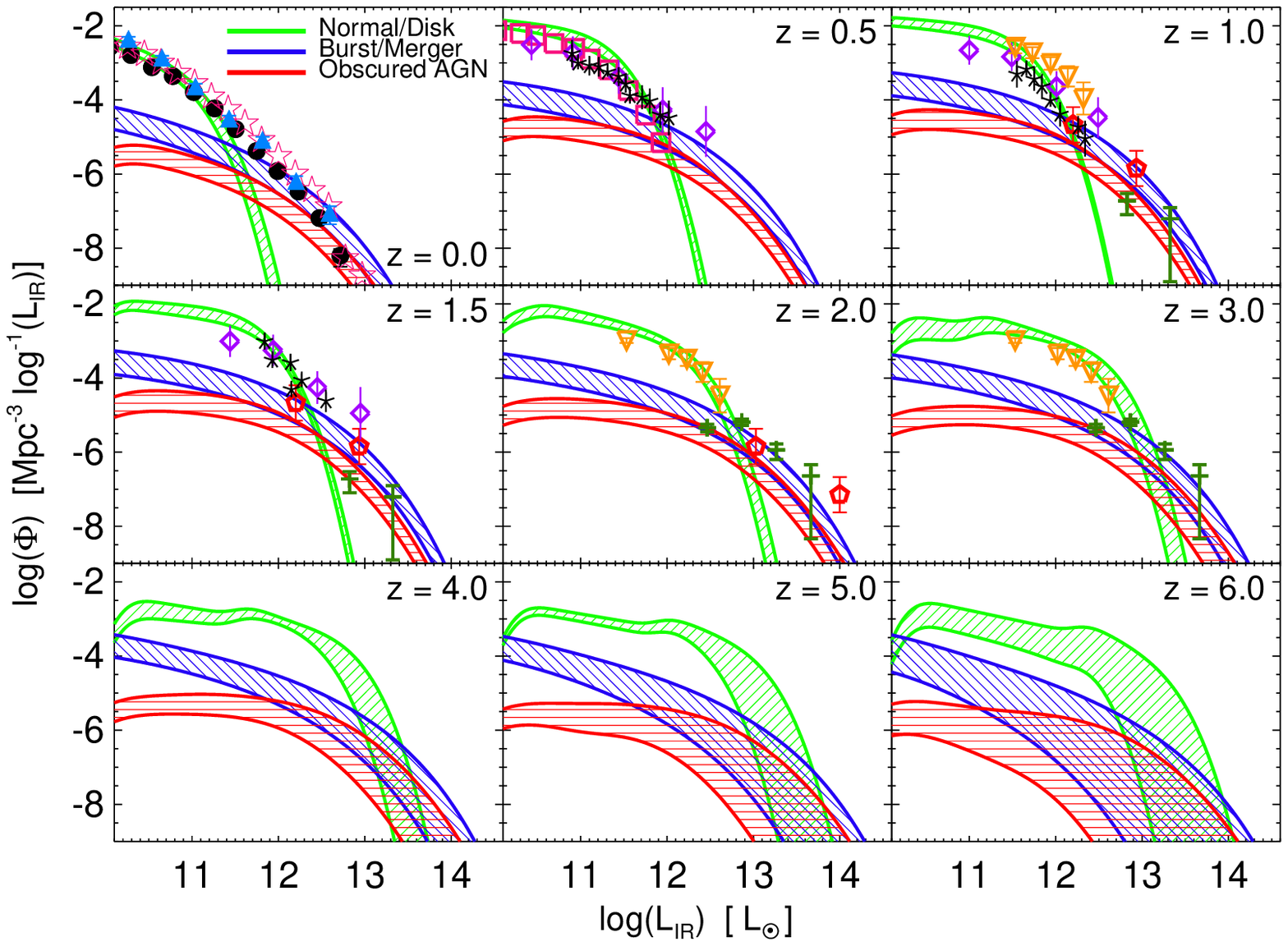}
    \caption{As Figure~\ref{fig:ir.lfs}, but with all calculations adoptng a steeper index in the 
    Kennicutt relation (Equation~\ref{eqn:kennicutt}), $n_{s}=1.6$, 
    normalized to the same SFR for Milky Way-like disks. Systems at high redshifts 
    are boosted significantly in SFR, and mergers are more concentrated 
    in time, leading to sharper peak SFRs. The agreement with observations 
    is somewhat improved, but overall the differences are comparable to the uncertainties 
    from the adopted SFR. 
    \label{fig:ir.lfs.steepks}}
\end{figure}

The consequences of adding, removing, or changing some of our model 
assumptions are discussed in \S~\ref{sec:lfs:tests}. Here, 
in Figures~\ref{fig:ir.lfs.nogasevol}, \ref{fig:ir.lfs.nosizeevol}, \ref{fig:ir.lfs.noscatter}, 
\&\ \ref{fig:ir.lfs.steepks}, we explicitly illustrate 
the effects discussed there.

In Figure~\ref{fig:ir.lfs.nogasevol} we reproduce Figure~\ref{fig:ir.lfs}, but 
do not allow galaxy gas fractions to evolve with redshift (adopting the $z=0$ 
value at all redshifts). As discussed in \S~\ref{sec:lfs:tests}, this leads to significant under-prediction 
of the IR LF at high redshifts. 
In Figure~\ref{fig:ir.lfs.nosizeevol}, we reproduce Figure~\ref{fig:ir.lfs} again, 
but this time do not allow for disk sizes to be more compact at high redshift. 
This has a much smaller effect. 
In Figure~\ref{fig:ir.lfs.noscatter}, we repeat Figure~\ref{fig:ir.lfs} but do not allow for 
scatter in any quantities (e.g.\ disk sizes, gas fractions, burst masses, quasar 
bolometric corrections, and SFRs at otherwise fixed properties); i.e.\ all 
values exactly trace the medians given in \S~\ref{sec:model}. 
This suppresses the bright end of the LF. 
In Figure~\ref{fig:ir.lfs.steepks}, we repeat Figure~\ref{fig:ir.lfs}, 
but adopt a steeper power-law index for the \citet{kennicutt98} relation 
($\dot{\Sigma}_{\ast}\propto\Sigma_{\rm gas}^{1.6}$), as discussed 
in \S~\ref{sec:lfs:tests}). These LFs correspond to the fits for the steep 
Kennicutt-Schmidt slope case presented in Table~\ref{tbl:fits}.

\end{appendix}

\end{document}